\documentstyle[aps,eqsecnum,epsf,tighten,preprint]{revtex}
\begin{document}
\widetext
\title{Theory of finite temperature crossovers near quantum critical points close to, or
above, their upper-critical dimension}
\author{Subir Sachdev}
\address{
Department of Physics, P.O. Box 208120,
Yale University, New Haven, CT 06520-8120}
\date{June 5, 1996}
\maketitle
\begin{abstract}
A systematic method for the computation of finite temperature ($T$) crossover functions near
quantum critical points close to, or
above, their upper-critical dimension is devised. 
We describe the physics of the various regions in the $T$ and critical tuning 
parameter ($t$) plane. The quantum critical point is at $T=0$, $t=0$, and in 
many cases there is a
line of finite temperature transitions at $T = T_c (t)$, $t < 0$ with $T_c (0) = 0$. For the
relativistic, $n$-component $\phi^4$ continuum 
quantum field theory (which describes lattice quantum rotor ($n \geq 2$) and transverse field
Ising ($n=1$) models) the upper critical dimension is $d=3$, and for $d<3$,  $\epsilon=3-d$ 
is the control parameter over the entire phase diagram. 
In the region $|T - T_c (t)| \ll T_c (t)$, we obtain an $\epsilon$ expansion 
for coupling constants
which then are input as arguments of known {\em classical, tricritical,} crossover functions. 
In the high $T$ region of the continuum theory, an expansion in integer powers of
$\sqrt{\epsilon}$, modulo powers of $\ln \epsilon$, holds for all thermodynamic observables,
static correlators, and dynamic properties at all Matsubara frequencies; for the imaginary part
of correlators at real frequencies ($\omega$), the perturbative $\sqrt{\epsilon}$ expansion
describes quantum relaxation at $\hbar \omega \sim k_B T$ or larger, but fails for $\hbar \omega
\sim \sqrt{\epsilon} k_B T$ or smaller. An important principle, underlying the whole
calculation, is the analyticity of all observables as functions of $t$ at $t=0$, for $T>0$;
indeed, analytic continuation in $t$ is used to obtain results in a portion of the phase
diagram. Our method
also applies to a large class of other quantum critical points and their associated continuum
quantum field theories.
\end{abstract}
\pacs{PACS numbers:}
\widetext
\newpage
\section{Introduction}
\label{intro}
The study of finite temperature crossovers in the vicinity of quantum phase
transitions is a subject with a long
history~\cite{hertz,suzuki,lawrie,otherbose,cres,weichmann,dsf,CHN,CSY,millis,sokolpines,sss,contin,conserve,uz,scs,ioffe,qrsg,oppermann,anirvan,statphys}, 
but many aspects of it remain poorly understood. The structure of the crossovers is especially
rich for the case where the quantum critical point extends into a line
of finite temperature phase transitions,  and there is a reasonable qualitative 
understanding of all the regimes. While there have been quantitative calculations of crossover
functions in special cases~\cite{weichmann,CSY,scs,qrsg,oppermann,anirvan} there is no complete,
general theory of these crossovers, especially for the case when the quantum critical point is
below its upper critical dimension.

In this paper, we shall provide a new systematic and controlled approach to the quantitative
computation of these crossover functions. Our method is quite general: 
it should apply to
essentially all quantum critical points in the vicinity of, or above, their
upper-critical dimension. 

Recently, O'Connor and Stephens~\cite{oconnor} have also studied crossovers near
relativistic  quantum-critical points below their upper-critical dimension. They found it
necessary to introduce a non-standard extension of the field-theoretic renormalization group. We
will comment on their results (and of others) in Section~\ref{conc}. 

In this paper, we will show that it is possible to devise a simple strategy,
completely with the framework of standard field-theoretic methods, which provides a
systematic computation of the required crossovers. We shall describe how our method can be
extended to arbitrary orders in an expansion in powers of the interactions,
but we shall only provide here explicit computations at low orders. 
One of the main virtues of our method is that it clearly separates contributions of
fluctuations of different physical origins: critical singularities of the $T=0$
quantum-critical point, and those of the finite $T$ classical phase transition, are
accounted for at distinct stages of the calculation. 

We shall present most of our discussion in the context of a continuum quantum
field theory (CQFT) of a $n$-component bosonic field $\phi_{\alpha}$ ($\alpha = 1 \ldots
n$; we will drop the index $\alpha$ except where needed) with $O(n)$ symmetry and with
the bare, imaginary time ($\tau$) action
\begin{equation}
{\cal S} = \int_0^{1/T} d\tau \int d^d x \left\{ \frac{1}{2} \left[
(\partial_{\tau} \phi)^2 + (\nabla \phi)^2 + (m_{0c}^2 + t_0) \phi^2 \right]
+ \frac{u_0}{4!} \phi^4 \right\}.
\label{cals}
\end{equation}
We have set $\hbar = k_B = 1$, measured length scales ($x$) in units
in which the velocity of excitations $c=1$, and introduced the bare mass
$m_{oc}^2 + t_0$ and the bare coupling $u_0$ (the mass term has been separated so
that the
$T=0$ quantum critical point is at $t_0 = 0$). This field theory describes the
low-energy physics in the vicinity of the quantum phase transition in the
$d$-dimensional transverse-field Ising model (for $n=1$) or the $O(n)$ quantum
rotor model (for
$n>1$). The generalization of our method to other quantum field theories,
like the dilute Bose gas, or the models for onset of antiferromagnetism in Fermi
liquids, is straightforward and will also be discussed.

We begin our discussion by reviewing the expecting scaling structure of ${\cal S}$
for the case where the quantum critical point is below its upper critical
dimension. At
$T=0$,
${\cal S}$ describes the usual  $\phi^4$ theory in $d+1$ dimensions, and its upper
critical dimension is
$d=3$; for $d<3$, there is an essentially complete understanding~\cite{bgz,zj} of the critical
properties of this theory in an expansion in powers of
\begin{equation}
\epsilon = 3 - d.
\end{equation}
The definition of the renormalized theory requires a field scale renormalization
$Z$, a coupling constant renormalization $Z_4$, and a renormalization of $\phi^2$
insertions in the critical theory $Z_2$. In terms of these, we define us usual
\begin{equation}
t =t_0 Z / Z_2;
\label{deft}
\end{equation}
$t$ is a measure of the deviation of the system from its $T=0$ quantum critical
point. Precisely the same renormalizations are also sufficient to define a finite
theory at non-zero $T$, {\em even in the vicinity of the finite $T$ phase
transition line\/}, as we shall explicitly see in this paper. 

\begin{figure}
\epsfxsize=14in
\centerline{\epsffile{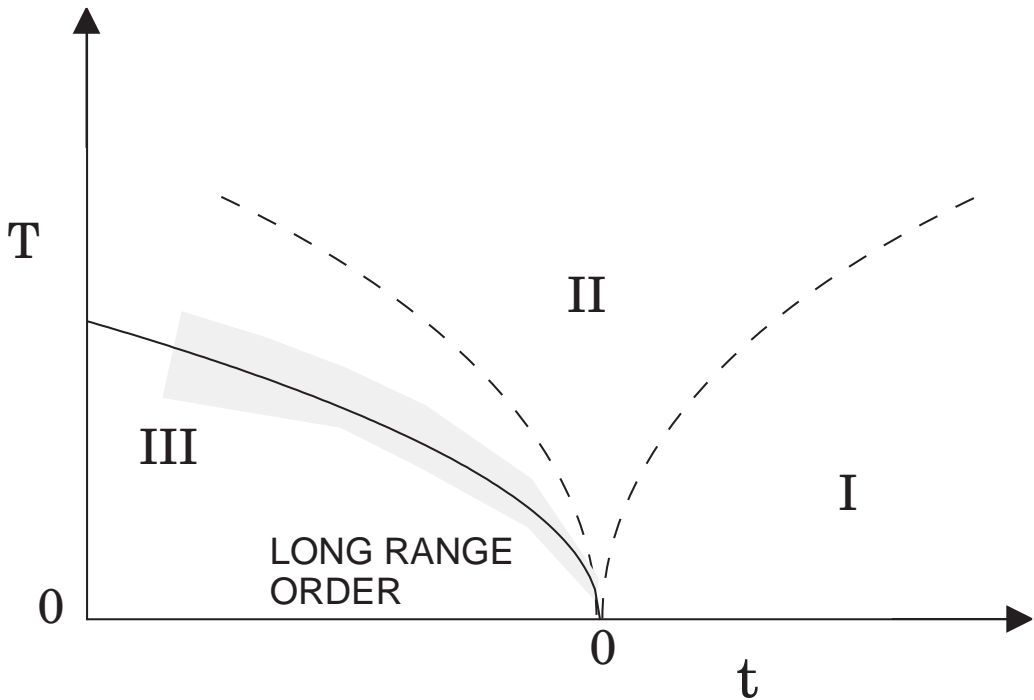}}
\vspace{0.5in}
\caption{
Finite temperature ($T$) phase diagram, as a function of the tuning parameter $t$, of crossovers near
a quantum critical point ($t=0$, $T=0$) below its upper critical dimension, for the case where
long-range-order survives at non-zero $T$. 
The dashed lines indicate smooth crossovers, while the full
line is the locus of finite temperature transitions $T= T_c (t)$. 
All crossover and phase-transition boundaries scale as $T \sim |t|^{z \nu}$, where $z$ is the
dynamic exponent and $\nu$ is the correlation length exponent. For the action ${\cal S}$ (Eqn
(\protect\ref{cals})) this paper provides a systematic expansion of observables with
$\epsilon=3-d$ the control parameter. In region II the expansion is in powers of
$\protect\sqrt{\epsilon}$ ($\epsilon = 3-d$), with additional factors of $\ln \epsilon$; in the
shaded region (defined by $|T - T_c (t)| \ll T_c (t)$) we provide an $\epsilon$-expansion of
coupling constants, which then  become arguments of previously known {\em classical,
tricritical,} crossover functions.  Elsewhere, all observables can be obtained in an expansion in
integer powers of $\epsilon$. For $T \neq 0$, 
all observables are analytic as a function of $t$ at $t=0$; in the region $t<0$, $T> T_c (t)$,
our results are obtained by analytic continuation from the $t>0$ results. 
All properties of the phase diagram are described by
the continuum quantum field theory associated with the quantum critical point,  which is in its
low $T$ limit in regions I and III, and in its high $T$ limit in region II. 
The condition determining boundary of region III is similar to that determining $T_c$ to within
constants of order unity that we are free to choose, and we have chosen region III to extend
to both sides of $T=T_c (t)$. For more discussion on the regions see
Section~\protect\ref{sec:susc} }
\label{f1}
\end{figure}

We show a sketch of the phase diagram of ${\cal S}$ as a function of $t$ and $T$
in Fig~\ref{f1}~\cite{statphys}. 
We have assumed in this figure, and throughout the remainder of the paper that 
the model ${\cal
S}$ is in its scaling limit {\em i.e.\/} the coupling $u_0$ is 
at its fixed point value, and all
ultraviolet cutoffs have been sent to infinity after an appropriate 
renormalization procedure.
There is a finite temperature phase transition line 
at $T=T_c (t)$, and all other boundaries are
smooth crossovers. All of the physics is contained within 
universal quantum-critical crossover
functions, which we now briefly describe. We will consider 
the behavior of the dynamic two-point
susceptibility, $\chi (k, \omega)$ ($k$ is a spatial momentum, 
and $\omega$ is a frequency)
obtained after analytic continuation of the susceptibility  \begin{equation}
\chi(k, i\omega_n ) =  \frac{1}{Z}\int d^d x \int_0^{1/T} d \tau 
e^{i(k x - \omega_n \tau)} \langle \phi (x,\tau) \phi (0,0) \rangle
\label{chidef}
\end{equation}
which is evaluated at Matsubara frequency $\omega_n = 2 n \pi T$. 
We will consider the scaling behavior of $\chi$ for $t>0$ and $t<0$ separately, and then
discuss the relationship between the two cases.

\noindent
({\em i}) $t>0$: 
The susceptibility $\chi$ obeys the scaling form~\cite{CSY}
\begin{equation}
\chi (k, \omega ) = {\cal A} \left( \frac{\hbar c}{k_B T} \right)^2
\left( \frac{k_B T}{\Delta_+} \right)^{\eta} \Psi_+
\left( \frac{\hbar c k }{k_B T}, \frac{\hbar \omega}{k_B T} , \frac{\Delta_+}{k_B T}
\right)~~~~~~~~t>0
\label{scalep}
\end{equation}
where we have momentarily re-inserted all factors of $\hbar$, $k_B$, and $c$,
$\eta$ is the usual field anomalous dimension of the $T=0$, $d+1$ dimensional theory,
and $\Psi_+$ is a fully universal, complex-valued, universal scaling function.
Notice that there are no arbitrary scale factors, and $\chi$ is fully determined by two
parameters, $\Delta_+$ and ${\cal A}$, which are properties of the $T=0$ theory.
The first of these, $\Delta_+$, is the true energy gap above the ground state, while the
second,
${\cal A}$, is the residue of the lowest quasi-particle excitation; they obey
\begin{equation}
{\Delta}_+ \sim t^{\nu}~~~~;~~~~ {\cal A} \sim t^{\eta \nu}~~~~~~~~~~t>0
\end{equation}
where $\nu$ is the usual correlation length exponent of 
the $T=0$ theory (all Greek letter
exponents in this paper refer to those of $T=0$ the 
quantum-critical point, and not to those the
finite $T$ phase transition). We provide a computation of the values of the
$T=0$ parameters,
$\Delta_+$ and
${\cal A}_+$, in Appendix~\ref{zeroT}.

The factors in front of $\Psi_+$ in (\ref{scalep}) have been chosen so that 
$\Psi_+$ is finite at $\Delta_+ / k_B T = 0$. All scaling functions defined in this
paper will share this property.

We also emphasize that although the scaling ansatz (\ref{scalep}) contains
dynamic information, its form and content are quite different from the 
dynamic scaling hypotheses applied near classical phase 
transitions~\cite{halphoh}. In these classical systems, a single diverging 
correlation length, $\xi$, is used to set the scale for $k$ and $\omega$; the
analog of (\ref{scalep}) is then a scaling function of {\em two \/}
arguments, $k \xi$ and $\omega \xi^{z_c}$, where $z_c$ is the classical dynamic
critical exponent. In contrast, the quantum crossover result (\ref{scalep})
is a function of {\em three\/} arguments, the extra argument arising
because the quantum critical point has two relevant perturbations
($T$ and $t$). Further, the identification of universal scale factors,
and indeed the conceptualization of the physics, is much more transparent
when $T$ is used as the primary energy setting the scale for other 
perturbations. Only in the immediate vicnity of the finite $T$ phase transition at $T_c (t)$,
$| T - T_c (t) | \ll T_c (t)$, does (\ref{scalep}) collapse into a scaling function
of two arguments, as has been discussed in Refs~\cite{CSY,ssising}.

\noindent 
({\em ii}) $t<0$:
Now the $T=0$ ground state breaks a symmetry with 
\begin{equation}
\langle \phi_{\alpha} \rangle
= N_0 \delta_{\alpha 1}~~~~~~~~T=0, t<0;
\end{equation}
here $N_0 \sim (-t)^{\beta}$ is the condensate, which we have arbitrarily chosen to
point in the $\alpha=1$ direction, and $\beta = (d-1+\eta)\nu/2$ is the magnetization
exponent of the $T=0$ theory. Now $N_0$ can serve as the parameter which determined the
field scale (replacing ${\cal A}$), and we need an energy scale which determines the
deviation of the ground state from the $t=0$, $T=0$ quantum critical point.
For $n=1$, there is a gap, $\Delta_-$, above the ground state, which satisfies our
requirements; we have therefore~\cite{CSY}
\begin{equation}
\chi (k, \omega ) = \frac{N_0^2 (\hbar c)^{d}}{\Delta_-^{d-1+\eta}
(k_B T)^{2-\eta}} \Psi_{-}
\left( \frac{\hbar c k }{k_B T}, \frac{\hbar \omega}{k_B T} , \frac{\Delta_-}{k_B T}
\right)~~~~~~~t<0, n=1.
\label{scalem1}
\end{equation}
For $n\geq 2$, there is no gap above the ground state, and we use instead the
spin-stiffness, $\rho_s \sim (-t)^{(d-1)\nu}$ as a measure of the deviation from the
quantum critical point; in this case we have the scaling form
\begin{equation}
\chi_{\alpha} (k, \omega ) = \frac{N_0^2}{\rho_s} \left( \frac{\hbar c}{k_B T} \right)^2 
\left( \frac{ (k_B T)^{d-1}}{ (\hbar c)^{d-2} \rho_s } \right)^{\eta/(d-1)}
\Psi_{-\alpha}
\left( \frac{\hbar c k }{k_B T}, \frac{\hbar \omega}{k_B T} , \frac{(\hbar c)^{d-2}
\rho_s}{(k_B T)^{d-1}}
\right)~~~~~t<0, n\geq 2.
\label{scalem2}
\end{equation}
A computation of the $T=0$
parameters $N_0$, $\Delta_-$ and $\rho_s$ is provided in Appendix~\ref{zeroT}.
The $\alpha$ dependence in (\ref{scalem2}) accounts for the difference between
fluctuations transverse and longitudinal to the condensate orientation.
Again, $\Psi_-$ is finite at $\Delta_- /k_B T = 0$, or at $\rho_s /T^{d-1} = 0$,
and all subsequent scaling functions will share this property.
The finite temperature phase transition at $T_c (t)$ is contained
entirely within the scaling function $\Psi_-$: this transition appears as 
a point of non-analyticity of $\Psi_-$ as a function of $\Delta_-/T$ or $\rho_s /
T^{d-1}$. An immediate consequence is that the value of $T_c$ can be determined
precisely in terms of the $T=0$ energy scale; we found, in an expansion in powers of
$\epsilon$, that
\begin{equation}
T_c = \frac{\Delta_-}{k_B \sqrt{\epsilon}} \left[ \frac{3}{2\pi} + {\cal O}
(\epsilon, \epsilon^{(1+\epsilon)/(1-\epsilon)}, \epsilon^{1/(2\nu )} )
\right]~~~~~~n=1
\label{restc1}
\end{equation}
and
\begin{equation}
T_c = \frac{ (\hbar c)^{(d-2)/(d-1)} \rho_s^{1/(d-1)}}{k_B} \left[ \left( \frac{3}{2
\pi^2 (n+2)}
\right)^{1/2}
 + {\cal O}
(\epsilon, \epsilon^{(1+\epsilon)/(1-\epsilon)}, \epsilon^{1/(2\nu )} )
\right]~~~~~~n\geq 2
\label{restc2}
\end{equation}
The various higher order contributions are all universal, but arise from very different
physical effects; we will discuss their origin later in the paper.
In the upper critical dimension ($\epsilon = 0$), these formulae are modified by
replacing $\epsilon$ by $1/\ln$'s : so for $n=1$, $T_c \sim \Delta_-  \ln^{1/2}
(1/\Delta_-)$ etc. 
For the cases $n=1$, $d=1$ and $n \geq 3$, $d=2$ it is known that in
fact $T_c = 0$ {\em i.e.} long-range order is present only at $T=0$, and disappears at any
non-zero $T$. For these cases, it is clear that the above results for $T_c$, and other
results obtained in the $\epsilon$ expansion, cannot be used in the region labeled III in
Fig~\ref{f1}. However, the results of this paper {\em can} still be usefully applied to
the remainder of the phase diagram of Fig~\ref{f1}. 

Although we have used two separate scaling forms to describe the behavior for $t<0$ and
$t>0$, there is a crucial connection between them. Notice that there is no
thermodynamic singularity at $t=0$ provided $T> 0$. This implies that the observable
$\chi(k, \omega )$ (and indeed all other observables) must be \underline{analytic as a
function of $t$ at $t=0$ as long as $T> 0$}. This principle will serve as an extremely
important constraint on the calculation in this paper; indeed, our method is designed
to ensure that analyticity holds at each order. Further, our results in the region $t<0$,
$T > T_c (t)$, were obtained by a process of analytic continuation from the $t>0$,
$T>0$ region (see Fig~\ref{f1}). The ground state for $t<0$ has a 
spontaneously broken symmetry,
and hence cannot be used to access the symmetric $t<0$, $T > T_c (t)$ region 
in perturbation
theory; instead it is more naturally accessed from the disordered side with $t>0$.
A similar procedure of analytic continuation in coupling constant was used recently in
exact determinations of quantum critical scaling functions in $d=1$~\cite{ssising}.

Before turning to a description of our method in Section~\ref{sec:method},
 we highlight one of
our new results. A particularly interesting property of CQFT's at finite $T$ is the expected
thermal relaxational behavior of their correlators in real time. This behavior cannot be
characterized simply to a field-theorist who merely considers correlators of ${\cal S}$,
defined as a CQFT in imaginary time with a spacetime geometry $R^d \times S_1$ (the tensor
product of infinite $d$-dimensional flat space with a circle of circumference $1/T$).
In real frequency, the thermal relaxational behavior is characterized by the fact that
$\lim_{\omega \rightarrow 0} \mbox{Im} \chi (k, \omega)/\omega$ is expected to be
finite, with the limiting value proportional to a relaxation constant. In
Ref~\cite{statphys}, the quantity
\begin{equation}
\Gamma_R^{-1} \equiv i \chi(0,0) \left. 
\frac{\partial \chi^{-1} (0, \omega)}{\partial \omega} \right|_{\omega = 0} 
\label{defgammar}
\end{equation}
was introduced as
a convenient characterization of the relaxation rate. By (\ref{scalep}), $\Gamma_R$ 
must obey a scaling form given by 
\begin{equation}
\Gamma_R = \frac{k_B T}{\hbar} \Psi_{\Gamma +} \left(
\frac{\Delta_{+}}{k_B T} \right),
\end{equation}
for $t>0$,
and similarly for $t<0$. In particular, in the high $T$ limit of the
CQFT~\cite{statphys} (this is region II of Fig~\ref{f1}), 
$\Gamma_R$ is $(k_B T/\hbar)$ times $\Psi_{\Gamma +} (0)$, which is 
expected to be a finite, universal number. 
Unfortunately, we shall find that the
perturbative expansion discussed in this paper cannot be used to obtain a systematic
expansion for $\Gamma_R$. A self-consistent approach, with damping of intermediate states,
appears necessary and will not be discussed here. In the high $T$ limit, we shall show that the
non-self-consistent approach fails for frequencies of order $\sqrt{\epsilon} T/\hbar $ or
smaller. To avoid this difficulty, let us define an alternative characterization of the damping
at frequencies of order $k_B T/\hbar$ by
\begin{equation}
\Gamma_{RT}^{-1} \equiv - \chi(0,0) \left. 
\mbox{Im} \frac{\chi^{-1} (0, \omega)}{\omega} \right|_{\omega = k_B T/\hbar}
\label{defgammat}
\end{equation}
This will obey a scaling form identical to that of $\Gamma_R$. We found in the high $T$ limit
that
\begin{equation}
\frac{k_B T}{\hbar \Gamma_{RT}} = \epsilon \frac{3}{(n+8) \pi}
\left[ \frac{1}{8} + \pi^2  + 6  \mbox{Li}_2 ( e^{-1/2}) \right] + {\cal O}(\epsilon^{3/2}),
\end{equation}
where $\mbox{Li}_2$ is the dilogarithm function defined in (\ref{defdilog}). The result will be
compared with an exact result for $\Gamma_{RT}$ for $d=1$, $n=1$ in Section~\ref{twoloop}.
Also in the exact result at $d=1$, $n=1$ we find $\Gamma_{RT}/\Gamma_{T} = 
0.98170018\ldots$ in the high $T$ limit, and we expect a similar ratio close to unity in all
cases in region II of Fig~\ref{f1}.

The following subsection contains a description of our approach. Readers not interested in the
details of its application to ${\cal S}$, can read Section~\ref{sec:method}, and skip ahead to
Section~\ref{conc} where we give our unified perspective of this and earlier work.

We will revert to setting
$\hbar = k_B = c = 1$ in the remainder of the paper.

\subsection{The Method}
\label{sec:method}
The origin of the approach we shall take here can be traced to early work by
Luscher~\cite{luscher} on the quantum $O(n)$ non-linear sigma model in $1+1$ dimensions.
Subsequently, a related idea was employed by Brezin and Zinn-Justin~\cite{bz} and by Rudnick,
Guo and Jasnow~\cite{rudnick} in their study of finite-size scaling crossover functions in
systems which are finite in all, or all but one, dimensions (also referred to as the
$d\rightarrow 0$ and
$d\rightarrow 1$ crossovers). The quantum-critical crossovers are clearly related, but now
involve
$d+1 \rightarrow d$. We shall show here that the latter problem can be successfully
analyzed by essentially the same method as that used for the former. There are some
new subtleties that arise in a limited region of the phase diagram, and we will
discuss below how they can be dealt with.

We will describe the method here for the special case of the action ${\cal S}$
(Eqn (\ref{cals})) with $d$ below its upper critical dimension {\em i.e.} $d<3$.
The central idea is that at finite $T$, it is safe to integrate out all modes of
$\phi (k, \omega_n )$ with a non-zero Matsubara frequency, $\omega_n \neq 0$, 
to derive an effective action for zero frequency modes $\phi(k, \omega_n = 0)$.
All modes being integrated out are regulated in the infrared by the $\omega_n^2$
term in their propagator, and so the process is necessarily free of infrared
divergences; further, the renormalizations of the $T=0$ theory also control the
ultraviolet divergences at finite $T$. To be specific, let us define
\begin{equation}
\Phi (x) \equiv \frac{T}{\sqrt{Z}} \int_0^{1/T} d \tau  \phi(x, \tau), 
\end{equation}
and its transform in momentum space $\Phi (k) = \int d^d x e^{ikx} \Phi(x)$.
Then from ${\cal S}$, we can deduce an effective action for $\Phi (k)$ after completely
integrating out the $\phi(k, \omega_n \neq 0)$ (we have set the coupling $u_0$ at the
fixed-point of its $\beta$-function---see Appendix~\ref{zeroT}):
\begin{eqnarray}
 && {\cal S}_{eff} = \frac{1}{T} \left[
\frac{1}{2}\int \frac{d^d k}{(2\pi)^d} \widetilde{C}_2 (k) |\Phi(k)|^2
\right.\nonumber\\ &&+ \left. \frac{1}{4!} \int \frac{d^d k_1 d^d k_2 d^d
k_3}{(2\pi)^{3d}} \widetilde{C}_4 (k_1 , k_2 , k_3,-k_1-k_2-k_3) \Phi (k_1 ) \Phi (k_2) \Phi
(k_3)
\Phi(-k_1-k_2-k_3) + \ldots \right].
\label{seff}
\end{eqnarray}
The couplings $\widetilde{C}_2$, $\widetilde{C}_4$, $\ldots$, are computed in a power series in
$\epsilon$, with the coupling constants renormalized as in the $T=0$, $d+1$ dimensional
critical theory. This procedure will remove all the ultraviolet divergences of the
quantum critical point. However, the ultraviolet divergences of the finite $T$,
$d$-dimensional $\phi^4$ theory remain; fortunately these are very simple as the
$\phi^4$ theory is {\em super-renormalizable\/} for $d < 4$~\cite{ramond}. In particular, $d<3$,
ultraviolet divergences are associated only with the one-loop `tadpole' graphs; so let
us define
\begin{equation}
C_2 (k) = \widetilde{C}_2 (k) + \left( \frac{n+2}{6} \right) \int \frac{d^d k_1 }{(2 \pi)^d}
\frac{\widetilde{C}_4 (k, -k, k_1, -k_1 )}{k_1^2} + \ldots
\label{tadpole}
\end{equation}
where the ellipses refer to ``tadpole'' contributions from higher order vertices like
$\widetilde{C}_6$, $\widetilde{C}_8$ $\ldots$. Similarly, there will also be tadpole
renormalizations of $\widetilde{C}_4$  to $C_4$ by higher order vertices, and so on.
These new vertices,  $C_{2p}$, $p$ integer, are now free of all ultraviolet 
divergences (for $3 \leq d < 4$ there is a second classical renormalization at the two-loop
level which must be accounted for; we will ignore this complication here and deal with it
later in the paper). They are also automatically free of infrared divergences as we are
only integrating out modes with a finite frequency. Indeed, these vertices must obey the
scaling forms:
\begin{equation}
C_{2p} (k_i ) = T^{d+1 - p(d-1+\eta)} \left(
\frac{\Delta_+^{\eta}}{{\cal A}}
\right)^p \Psi_{Q+}^{(p)} \left( \frac{k_i}{T}, \frac{\Delta_+}{T} \right),
\label{cscale}
\end{equation}
for $t>0$, with the $\Psi_{Q+}^{(p)}$ a set of quantum scaling functions; the subscript
$Q$ emphasizes that these scaling functions are properties of the $T=0$ quantum-critical
point, and distinguishes them from classical crossover functions we shall use later.
Similar scaling results hold for
$t<0$ and we will refrain from explicitly displaying them. 
For our subsequent discussion it is useful to define a set of couplings $K$, $R$,
and $U$, which can be obtained from the $C_{2p}$, and which play an important role in
our analysis:
\begin{eqnarray}
R &\equiv& C_2 (0) \nonumber \\
K &\equiv& \left( 1 + \left. \frac{\partial C_2 (k)}{\partial k^2}
\right|_{k=0}\right)
\nonumber \\
U &\equiv& C_4 (k_i = 0) 
\label{defu}
\end{eqnarray}
It is clear that $R$, $K$, and $U$ obey scaling forms that can be easily deduced from
(\ref{cscale}).

We also note that the
couplings $R$ and $C_{2p} (k_i)$ are guaranteed to be analytic as a function of $t$ at
$t=0$; this is because only finite frequency modes have been integrated out, and their
propagator $1/(k^2 +
\omega_n^2 + t)$ is not singular at $t=0$ in the infrared ($k=0$) as $\omega_n \neq  0$. 
This analyticity
will be of great use to us later.

Assume, for the rest of this section, that we know the $\Psi_{Q \pm}^{(p)}$ functions
(we will provide explicit computations of some of them later in the paper). We are now
faced with the seemingly difficult problem of computing observables in the $\Phi$ theory
with the action
${\cal S}_{eff}$. This is a theory in dimension $d$ close to 3 ({\em not} $d$ close to
4), and one might naively assume, that this problem is intractable. We shall now argue
that, in fact, it is not. The argument is contained in the following two simple, but
important, observations.\\ 
(1) Consider a perturbation theory of ${\cal S}_{eff}$ in
which the propagator is
$1/(K k^2 + R)$, and we expand in the non-linearities $C_{2p} (k_i)$, $p\geq
2$ in powers of $\epsilon$. 
This expansion differs from an ordinary expansion in powers of $\epsilon$ only in that
the mass $R$ and coupling $K$ itself contain corrections in powers of $\epsilon$;
alternatively one could also treat $K-1$ as an interaction, and work with the propagator
$1/(k^2 + R)$, but it is essential to keep the mass $R$ in the propagator. Such a
procedure is guaranteed to be finite in the ultraviolet. This follows immediately from
the statement that the renormalization of the
$T=0$ theory (which we carried out while obtaining
${\cal S}_{eff}$ from ${\cal S}$) is sufficient to remove ultraviolet divergences even at
non-zero $T$. In other words, the momentum dependencies in the $C_{2p} (k_i)$ must be
such that all ultraviolet divergences cancel out.\\
(2) The action ${\cal S}_{eff}$ is weakly coupled over the bulk of the phase diagram in
the $t, T$ plane, and so the procedure in (1) leads to accurate results for physical
observables. Only in the region
$|T - T_c (t) |
\ll T_c (t)$ (drawn shaded in Fig~\ref{f1}) is a more sophisticated analysis
necessary, which will be described momentarily.  
To verify this claim, consider the values of the low order couplings in ${\cal
S}_{eff}$ at $t=0$, but $T$ finite; we will find later that
\begin{eqnarray}
R &\sim& \epsilon T^{2-\eta} \nonumber \\
K  &\sim&  T^{-\eta}
\nonumber \\
U  &\sim& \epsilon T^{\epsilon - 2 \eta}
\label{valkru}
\end{eqnarray}
for $t=0$.
(For the present purpose, we can neglect  all $C_{2p}$ for $p>2$ as they are all of 
order $\epsilon^2$.)  A dimensionless measure of the strength of non-linearities in
${\cal S}_{eff}$ is
\begin{equation}
\frac{T U}{K^{d/2} R^{(4-d)/2}} \sim \epsilon^{(1-\epsilon)/2} \sim \sqrt{\epsilon}
\left( 1 - \frac{\epsilon \ln \epsilon}{2} \right) \ll
1~~~~~~~~\mbox{for
$t=0$}.
\label{ratio}
\end{equation}
The above dimensionless ratio is simply that appearing in the familiar Ginzburg
criterion~\cite{ma}. So a simple perturbative calculation is adequate for $t=0$.
For $t>0$, the behavior of perturbation theory can only improve as the mass $R$ becomes
larger, which decreases the value of the above dimensionless ratio; as a result the
perturbative calculation describes the crossover between the quantum-critical and
quantum-disordered regimes of Fig~\ref{f1}. For
$t<0$ the perturbation theory is initially adequate, but eventually becomes unreliable
in the region $|T-T_c (t)| \ll T_c (t)$.

As the above results contain some the key points which allowed the computations of this paper,
it is useful to reiterate them. We describe the nature of our expansion, excluding the
shaded region
$| T - T_c (t)| 
\ll T_c (t)$ of Fig~\ref{f1}. The first step is to obtain an expansion for the `mass' $R$ of
the $\omega_n = 0$ mode: we outlined above a procedure which yields a series in integer powers
of $\epsilon$. Then, generate an expansion for the physical observable of interest, temporarily
treating $R$ as a fixed constant independent of $\epsilon$; this will again be a series in
integer powers of $\epsilon$, but strong infrared fluctuations of the $\omega_n =0 $ mode
lead to a singular dependence of the latter series on $R$.
Finally, insert the former series for $R$ into the latter series
for the physical observable.  In the quantum-disordered region
(Fig~\ref{f1}), $R$ is of order unity, and the final result remains a series in integer powers
of $\epsilon$. However, in the quantum-critical region, $R$ is of order $\epsilon$ (Eqn
(\ref{valkru})) and the result, by (\ref{ratio}), is a series in integer powers of
$\sqrt{\epsilon}$, with possibly a finite number of powers of $\ln \epsilon$ multiplying the
terms.  It is important to note that the final series in powers of $\sqrt{\epsilon}$ is
obtained from the original series in powers of $\epsilon$ only by a local re-arrangement of
terms {\em i.e.} given all the terms up to a certain order in the $\epsilon$ series,
we can obtain all terms below a related order in the $\sqrt{\epsilon}$ series.

Finally, let us turn our attention to the troublesome region $|T-T_c (t)| \ll T_c (t)$.
We expect this region to be dominated by the classical fluctuations characteristic of
the finite temperature transition, and hence to be well described by the following
action ${\cal S}_{C}$, which is a truncated form of ${\cal S}_{eff}$:
\begin{equation}
{\cal S}_{C} = \frac{1}{T} \int d^d x \left[
\frac{K}{2} ( \nabla {\Phi} )^2  + \frac{R}{2}
{\Phi}^2 +
\frac{ U}{4!}
{\Phi}^4
\right].
\label{scl}
\end{equation}
The couplings above were defined in (\ref{defu}). We have implicitly performed tadpole
renormalizations where necessary to remove ultraviolet divergences of the classical
theory. An immediate consequence of the super-renormalizability of the classical ${\cal
S}_C$ is that all observables are universal functions of the ``bare'' coupling
constants
$K$,
$R$ and $ U$. So for example, we have for the static susceptibility 
\begin{equation}
\chi(k , i\omega_n = 0) = \frac{1}{T} \langle | \Phi (k) |^2 \rangle
= \frac{1}{R} \Psi_{C} \left( \frac{Kk^2}{R}, \frac{T U}{K^{d/2} R^{(4-d)/2}} \right)
\label{psicl}
\end{equation}
where $\Psi_{C}$ is a universal crossover function with no arbitrary scale factors.
In fact, the crossover function $\Psi_{C}$ has been considered earlier in Ref~\cite{nr}, where it
was dubbed the {\em tricritical crossover function} for entirely different physical reasons (we
emphasize that this terminology is purely accidental - we are not dealing with any tricritical
point here). The computation of tricritical crossover functions is a logically separate problem
from those considered in this paper, and we shall have relatively little to say about them here.
We shall simply treat them as known, previously computed functions; for completeness, we tabulate
some results on these functions in Appendix~\ref{tricritical}. Notice that the arguments of the
classical crossover function
$\Psi_{C}$ in (\ref{psicl}) are themselves quantum-critical crossover functions, as follows from
(\ref{cscale}) and (\ref{defu}). Indeed, inserting (\ref{cscale}) and (\ref{defu}) into
(\ref{psicl}), we get a scaling form completely consistent with (\ref{scalep}).
The critical temperature, $T_c$, is determined by the condition $\chi(0,0) = \infty$;
in general, this is not equivalent to the requirement $R=0$ (although this does turn out to be
the case at the one-loop level), but is instead given by the point where the scaling function
$\Psi_C$ diverges. This condition leads to equation $TU/(K^{d/2} R^{(4-d)/2}) = \mbox{constant}$
where the constant is determined by the point where $\Psi_C$ diverges as a function of its
second argument; it is the solution of this equation which leads to the ${\cal
O}(\epsilon^{(1+\epsilon)/(1-\epsilon)})$ corrections to the result for $T_c$ reported in
(\ref{restc1}).

To summarize, in the region $|T - T_c (t) | \ll T_c (t)$, the physics is described by
universal crossover functions, $\Psi$, which are
``crossover functions ($\Psi_C$) of crossover functions ($\Psi_Q$)''. The $\Psi_Q$
functions are properties of the quantum-critical point, and it is the burden of this paper to
compute them; these functions then serve as arguments of known classical, tricritical crossover
functions ($\Psi_C$).

Finally, we note that for $d$ just below 3, it is also necessary to include a coupling
$V \Phi^6$ in ${\cal S}$ to get the correct infrared behavior; we have ignored this
complication for simplicity; moreover, as $V \sim \epsilon^3$ and this effect is
present only at a rather high order.

The outline of the remainder of the paper is as follows. In Section~\ref{sec:below} we will
compute properties of ${\cal S}$ for $d<3$. The discussion is divided into a one-loop
computation of static observables in Section~\ref{oneloop} and a two loop computation of
dynamic observables in Section~\ref{twoloop}. Section~\ref{sec:above} will present a general
discussion of the properties of models above their upper critical dimension: the modifications
necessary in the scaling forms and the explicit computation of crossover functions.
Finally Section~\ref{conc} will review the main results, discuss their relationship to previous
works, and point out directions for future work. A number of details of the calculations appear
in the appendices. The tricritical crossover functions appearing in (\ref{psicl}) are in
Appendix~\ref{tricritical}. In Appendix~\ref{zeroT} we compute the $T=0$ parameters that appear
in the scaling forms. Details of the finite $T$ two-loop computations of various quantities are
in Appendix~\ref{Tgt0}. 

\section{Crossover functions of ${\cal S}$ below three dimensions}
\label{sec:below}

A number of crossover functions for the model ${\cal S}$ were introduced
in Section~\ref{intro}. We give formal expressions, valid to two-loop order, for
all of these quantities in Appendix~\ref{Tgt0}. In this section we will evaluate
these expressions and show that they obey the required scaling forms order-by-order
in $\epsilon$. We will discuss the behavior for general values  of $t$ only 
to one-loop order in Section~\ref{oneloop}. We will limit our explicit two-loop
results to a few important quantities at the critical coupling $t=0$; these will appear
in Section~\ref{twoloop}.

The same basic trick will be repeatedly used to evaluate the frequency summations
in Appendix~\ref{Tgt0}: we will always subtract from the summation of a function
of $\omega_n$, the integration over frequency of precisely the same function. The 
resulting difference will always turn out to be strongly convergent as a 
function of momentum in all $d$. So, for example, we have:
\begin{eqnarray}
T\sum_{\epsilon_n} \int \frac{d^d q}{(2 \pi)^d}\frac{1}{\epsilon_n^2 + q^2 + a^2}
- && \int \frac{d\epsilon}{2 \pi} \int \frac{d^d q}{(2 \pi)^d} \frac{1}{\epsilon^2
+ q^2 + a^2}  \nonumber \\
&&=
T\sum_{\epsilon_n} \int \frac{d^d q}{(2 \pi)^d}
\frac{1}{\epsilon_n^2 + q^2 + a^2} - \int \frac{d^{d+1} p}{(2 \pi)^d}
\frac{1}{p^2 + a^2} \nonumber \\
&&= \int \frac{d^d q}{(2 \pi)^d} \frac{1}{\sqrt{q^2 + a^2}}
\frac{1}{e^{\sqrt{q^2+a^2}/T}-1}
\label{identity}
\end{eqnarray}
Notice that the integrand on the right hand side of the last equation falls off
exponentially for large
$q$, and the integral is therefore convergent for all $d$.
\subsection{One loop results}
\label{oneloop}
We will begin (Section~\ref{sec:cc}) by determining the coupling constants $R$, $K$ and $U$ of
${\cal S}_{cl}$ which are the arguments of
the tricritical crossover functions.
Subsequently, we will present results for observables of ${\cal S}$: 
the susceptibility $\chi$ (Section~\ref{sec:susc}) 
and the response of the system to a field that
couples to the conserved $O(n)$ charge (Section~\ref{sec:conserve}).

\subsubsection{Coupling constants of ${\cal S}_{cl}$}
\label{sec:cc}

We begin by using (\ref{resC2}) and (\ref{defu}) to obtain an expression for $R$,
valid to one-loop order:
\begin{equation}
R = t_0 + u_0 \left( \frac{n+2}{6} \right) \left[
\int \frac{d^d q}{(2 \pi)^d} \left( T \sum_{\epsilon_n \neq 0} \frac{1}{\epsilon_n^2 + q^2 +
t_0} + \frac{T}{q^2}\right)  - \int \frac{d^{d+1} p }{(2 \pi)^{d+1}} \frac{1}{p^2} \right]
\label{resR}
\end{equation}
We have set $Z=1$ at this order, and will implicitly do so in the remainder of this section.
We now apply the identity (\ref{identity}), perform the momentum integrals over $p$,
express in terms of the renormalized $t$ using (\ref{deft}). Finally we
express $u_0$ is terms of a renormalized coupling $g$ defined by 
\begin{equation}
g = \mu^{-\epsilon} S_{d+1} \frac{Z^2}{Z_4} u_0,
\label{defg}
\end{equation}
where $\mu$ is a renormalization momentum scale, $S_d = 2 /(\Gamma(d/2) (4
\pi)^{d/2})$ is a phase space factor, and the values of the renormalization constants are
tabulated in (\ref{defzs}). This gives for $t>0$:
\begin{eqnarray}
R =&& t \left( 1 + \frac{n+2}{6\epsilon} g \right) + \mu^{\epsilon} g 
\left( \frac{n+2}{12} \right) \Gamma \left( 2 - \frac{\epsilon}{2} \right)
\Gamma \left( -1 + \frac{\epsilon}{2} \right) t^{1 - \epsilon/2} \nonumber \\
&& ~~~~~~~~~~~~~~~~~~~~~~~~~~~~~~+ g  \left( 
\frac{n+2}{6} \right) T^2 \left(\frac{\mu}{T} \right)^{\epsilon} F_d \left( \frac{t}{T^2} \right) 
\nonumber \\ 
=&& t \left( 1 + \epsilon \frac{n+2}{2(n+8)} \ln \frac{t}{\mu^2} \right)
+ \epsilon T^2 \left( \frac{n+2}{n+8} \right) F_3 \left( \frac{t}{T^2} \right).
\label{resRtgt0}
\end{eqnarray}
In the second equation, we have evaluated at the fixed point value $g=g^{\ast}$ (Eqn
(\ref{valgast})), and then expanded to order $\epsilon$. The function
$F_d$ is given by
\begin{equation}
F_d (y) = \frac{1}{S_{d+1}}
\int \frac{d^d k}{(2 \pi)^d} \left[
\frac{1}{\sqrt{k^2 + y}}
\frac{1}{e^{\sqrt{k^2+y}}-1} - \frac{1}{k^2+y} + \frac{1}{k^2} \right].
\label{deffd}
\end{equation}
Let us also note here the values
\begin{equation}
F_d (0) =  \frac{(4 \pi)^{1/2} \Gamma((d+1)/2) \Gamma(d-1) \zeta (d-1)}{
\Gamma(d/2)}~~~~~~,~~~~~~F_3 (0) = \frac{2 \pi^2}{3}
\label{valfd}
\end{equation}
It is now easy to see, using (\ref{valdelp}) that the result (\ref{resRtgt0}) for $R$ can
be written in the scaling form
\begin{equation}
R = \frac{T^{2-\eta} \Delta_+^{\eta}}{{\cal A}} \left( 
\frac{\Delta_+^2}{T^2} + \epsilon  \left( \frac{n+2}{n+8} \right) F_3 \left(
\frac{\Delta_+^2}{T^2}
\right)\right).
\label{resRa}
\end{equation}
This result is consistent with the scaling postulated in (\ref{cscale}).
At this order, the exponent $\eta = 0 $, and verifying the powers of $\eta$ in front
requires a higher order computation. 

\begin{figure}
\epsfxsize=6in
\centerline{\epsffile{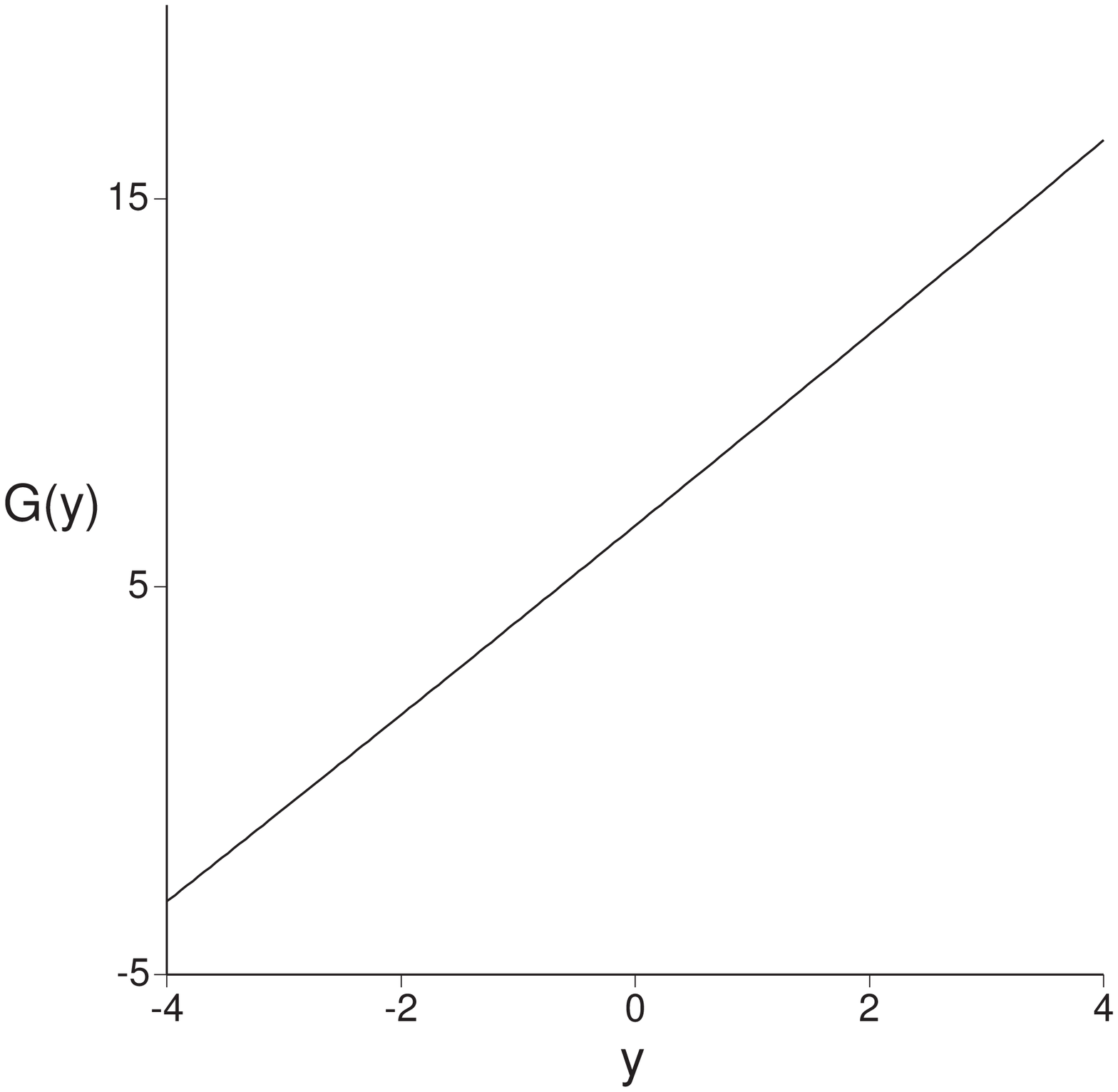}}
\vspace{0.5in}
\caption{A plot of the universal scaling function $G(y)$ defined in (\protect\ref{resG4}). Notice
that it is smooth at $y=0$. It is analytic for all real
$y > -2 \pi$. }
\label{f2}
\end{figure}

We now wish to extend this result for $R$ to $t< 0$ by analytic continuation from the
$t>0$ result. First, we need to verify that the $t>0$ result for $R$ is analytic at
$t=0$. To do this, we first rewrite (\ref{resRtgt0}) in the form 
\begin{equation}
R = t \left( 1 + \epsilon \frac{n+2}{n+8} \ln \frac{T}{\mu} \right) + \epsilon T^2
\frac{n+2}{n+8} G_+ \left( \frac{t}{T^2} \right)
\label{resR2}
\end{equation}
where
\begin{equation}
G_+ (y) = \frac{y \ln y}{2} + 4 \int_0^{\infty} k^{2} dk
\frac{1}{\sqrt{k^2 + y}}
\frac{1}{e^{\sqrt{k^2+y}}-1} + 2 \pi \sqrt{y}
\label{resG1}
\end{equation}
The first term in (\ref{resR2}) is clearly a smooth function of $t$; if we can now show
that $G_+ (y)$ is a smooth function of $y$ near $y=0$, we will have established the
analyticity of $R$ at $t=0$.
Performing an integration by parts of the integral in (\ref{resG1}), followed by
an elementary re-arrangement of terms, we can manipulate the result for $G_+ (y)$ into
the following form:
\begin{eqnarray}
G_+ (y) = && \frac{y \ln y}{2} + 2 \pi \sqrt{y}
- 4 \int_0^{\infty}  dk \ln \left[
1 - e^{-\sqrt{k^2+y}} \right] \nonumber \\
= && \frac{y \ln y}{2} + 2 \pi \sqrt{y} + 2 \int_0^{\infty} dk \left(
\sqrt{k^2 + y} - k - \ln \left[\frac{k^2+y}{k^2}\right] - \frac{y}{2\sqrt{k^2+1}} \right) 
\nonumber \\
&&~~~~~~~~-4 \int_0^{\infty} dk \left( \ln \left[
k \frac{\sinh(\sqrt{k^2+y}/2)}{\sqrt{k^2+y}/2} \right] - \frac{k}{2} -
\frac{y}{4\sqrt{k^2+1}} \right)
\end{eqnarray}
The first integral can be done analytically, and we find that all the potentially
singular terms cancel. Our final expression for $G_+ (y)$, valid for $y>0$ is:
\begin{equation}
G_+ (y) = \frac{y}{2} -4 \int_0^{\infty} dk \left( \ln \left[
k \frac{\sinh(\sqrt{k^2+y}/2)}{\sqrt{k^2+y}/2} \right] - \frac{k}{2} -
\frac{y}{4\sqrt{k^2+1}} \right)
\label{resG2}
\end{equation}
It should now be evident that (\ref{resG2}) is a smooth function of $y$ at $y=0$;
the integrand involves only even powers of $\sqrt{k^2 + y}$, and its integral is a
smooth function of $y$. Indeed, it is not difficult to explicitly extend
the above result to $z=-y < 0$. Divide the integral into the regions $k < \sqrt{z}$
and $k > \sqrt{z}$; the integrand remains unchanged in the second region, while in the
first region the $\sinh$ function becomes a $\sin$ function---this gives us
the function $G_- (z)$ as the analytic continuation of $G_+ (y)$ to $z=-y < 0$:
\begin{eqnarray}
G_- (z) = && \frac{z}{2} - z \sinh^{-1} \sqrt{z} -4 \int_0^{\sqrt{z}} dk  \ln \left[
k \frac{\sin(\sqrt{z-k^2}/2)}{\sqrt{z-k^2}/2} \right] \nonumber \\
&&~~~~~~~~~~~~~~~~
-4 \int_{\sqrt{z}}^{\infty} dk \left( \ln \left[
k \frac{\sinh(\sqrt{k^2-z}/2)}{\sqrt{k^2-z}/2} \right] - \frac{k}{2} +
\frac{z}{4\sqrt{k^2+1}} \right)
\label{resG3}
\end{eqnarray}
We can now combine the above results, to obtain an expression for $R$ valid for
both signs of $t$, and which is smooth at $t=0$:
\begin{eqnarray}
R &=& t \left( 1 + \epsilon \frac{n+2}{n+8} \ln \frac{T}{\mu} \right) + \epsilon T^2
\frac{n+2}{n+8} G \left( \frac{t}{T^2} \right) \nonumber \\
G(y) &\equiv& \theta(y) G_+ (y) + \theta(-y) G_- (-y)
\label{resG4}
\end{eqnarray}
A plot of the function $G(y)$ is shown in Fig~\ref{f2}. As expected, the plot is smooth at $y=0$.
However, the alert reader will notice that there is in fact a logarithmic singularity in $G$ ($R$)
at $y=-2\pi$ ($t = - 4\pi^2 T^2$) where the argument of the logarithm in (\ref{resG3}) can first
change sign. However, this singularity is of no physical consequence as it occurs
when the system is already in the ordered phase (Fig~\ref{f1}), and the above
expressions can no longer be used; the transition to  the ordered phase happens when $t
\sim - \epsilon T^2$. More precisely, we see from the tricritical function in
Appendix~\ref{tricritical} that the value of
$T_c (t)$ is determined by the condition $R=0$; applying this to (\ref{resG3}) and using
(\ref{valdelm}), (\ref{valn0}), (\ref{valrhos}) we get the results (\ref{restc1}),
(\ref{restc2}) for
$T_c$. We will discuss the physical significance of the limiting behavior of $R$ and $G$ in
various regimes in Section~\ref{sec:susc}.

Next, we turn to the computation of $U$. First,
we obtain from  (\ref{resC4}) and (\ref{defu}) the expression
\begin{equation}
U = u_0 - u_0^2 \left( \frac{n+8}{6} \right) T \sum_{\epsilon_n \neq 0}
\int \frac{d^d q}{(2 \pi)^d}  \frac{1}{(\epsilon_n^2 + q^2 + t_0)^2}
\label{resU}
\end{equation}
As in the computation of $R$, we can set $Z=1$, $\eta=0$ in the one-loop approximation.
Now note that
\begin{eqnarray}
T \sum_{\epsilon_n \neq 0}
\int \frac{d^d q}{(2 \pi)^d}  \frac{1}{(\epsilon_n^2 + q^2 + t_0)^2} = &&
\left( T \sum_{\epsilon_n}
\int \frac{d^d q}{(2 \pi)^d} \frac{1}{(\epsilon_n^2 + q^2 + t_0)^2} - \int
\frac{d^{d+1}p}{(2 \pi)^{d+1}} \frac{1}{(p^2 + t_0 )^2} \right) \nonumber \\
&&~~~~~~~- \left( T \int \frac{d^d q}{(2\pi)^d} \frac{1}{(q^2 +
t_0)^2} -
\int \frac{d^{d+1}p}{(2 \pi)^{d+1}} \frac{1}{(p^2 + t_0 )^2} \right) \nonumber \\
= && - T^{-\epsilon} \frac{d}{dy} \left( \int \frac{d^d q}{(2 \pi)^d} \frac{1}{\sqrt{q^2 + y}}
\frac{1}{e^{\sqrt{k^2 + y}} - 1} \right) \nonumber \\
&&~~~~~~~- T^{-\epsilon} S_{d+1} \left( - \frac{1}{\epsilon} + \frac{\pi}{\sqrt{y}}
+ \frac{\ln y + 1}{2} + {\cal O}(\epsilon) \right), 
\end{eqnarray}
where $y \equiv t_0/T^2$. Using the definition of the function
$G_+$ in (\ref{resG2}), we get finally
\begin{equation}
T \sum_{\epsilon_n \neq 0}
\int \frac{d^d q}{(2 \pi)^d}  \frac{1}{(\epsilon_n^2 + q^2 + t_0)^2} =
T^{-\epsilon} S_{d+1} \left( \frac{1}{\epsilon} - G^{\prime} \left( \frac{t_0}{T^2} \right)
+ {\cal O}(\epsilon) \right) 
\label{id1}
\end{equation}
Note that we have analytically continued $G'_+$ to $G'$ and obtained an expression which
is manifestly analytic at $t=0$. Inserting (\ref{id1}) into (\ref{resU}), expressing
$u_0$ and
$t$ in terms of the renormalized $g$ (Eqn (\ref{defg})) and $t$ (Eqn (\ref{deft})); this
yields
\begin{equation}
U = \mu^{\epsilon} g \left( 1 + \frac{n+8}{6\epsilon} g\right)
+ \left( \frac{n+8}{6\epsilon} \right) \frac{\mu^{2\epsilon} g^2}{T^{\epsilon}}
\left( -\frac{1}{\epsilon} + G' \left( \frac{t}{T^2} \right) \right)
\end{equation}
 Evaluating at $g=g^{\ast}$ (Eqn (\ref{valgast})) and
expanding to order
$\epsilon^2$, we get finally
\begin{equation}
U = \frac{6 \epsilon T^{\epsilon}}{n+8} \left[
1 + \epsilon \left( \frac{3 (3n+14)}{(n+8)^2} - \frac{1}{2} \right)
+ \epsilon G' \left( \frac{t}{T^2} \right) \right]
\end{equation}
Note that the $\mu$ dependence has dropped out at this order, and this result is
consistent with the scaling form (\ref{cscale}) and (\ref{defu}).

Finally, it is clear that the coupling $K=1$ at one loop.
 
\subsubsection{Susceptibility}
\label{sec:susc}

The one-loop susceptibility follows immediately from the result (\ref{reschi}):
\begin{equation}
\chi^{-1} (k, i\omega_n) = k^2 + \omega_n^2 + R - \epsilon 
\left(\frac{n+2}{n+8} \right) 2 \pi T \sqrt{R}
\label{reschioneloop}
\end{equation}
As $R$ is analytic at $t=0$, so is this result for $\chi$. 
It is also clear that this result obeys the scaling forms (\ref{scalep}),
(\ref{scalem1}), and (\ref{scalem2}), given that the result (\ref{resRa}) for $R$
obeys (\ref{cscale}).

The result (\ref{reschioneloop}) gives us a prediction for the $T$ and $t$ dependence
of the correlation length $\xi$:
\begin{equation}
\xi^{-2} = R - \epsilon 
\left(\frac{n+2}{n+8} \right) 2 \pi T \sqrt{R}
\label{resxi}
\end{equation}
By examining the
limiting behavior of $\xi$, we can obtain a physical interpretation of the regimes of the
CQFT associated with the $t=0$, $T=0$ quantum-critical point,
as shown in Fig~\ref{f1}.
\newline
(I) $T \ll \Delta_{+}$: low $T$ limit of CQFT; paramagnetic phase
\newline
From (\ref{resxi}) and (\ref{resRa}) we find
\begin{equation}
\xi^{-2} = \Delta_{+}^2 + \epsilon \left( \frac{n+2}{n+8} \right) T (8 \pi T \Delta_+ )^{1/2}
e^{-\Delta_+ / T} + \ldots
\end{equation}
So the correlation length, and the physics, is dominated by its $t>0$, $T=0$ behavior, with
exponentially small corrections due to a dilute number of thermally excited quasiparticles.
\newline
(II) $ T \gg |t|^{z\nu}$: high $T$ limit of CQFT
\newline
In this case, the leading behavior of $\xi$ from (\ref{resxi}) and (\ref{resRa}) is
\begin{equation}
\xi^{-2} = \epsilon \left(\frac{n+2}{n+8} \right)
 \frac{2 \pi^2 T^2}{3} \left[ 1 - \left( \frac{6 \epsilon (n+2)}{n+8} \right)^{1/2}
\right] 
\end{equation}
The scale of $\xi$, and indeed  of all the physics, is now set by $T$.
The ratio $\xi^{-2} / T^2$ is a universal number, obtained above for small $\epsilon$.
The reasons for the appearance of the $\sqrt{\epsilon}$ terms were discussed earlier in
Section~\ref{sec:method}; as also noted there, notice 
that there were no such terms in the region I.
This series for $\xi$ is not useful as it stands, as it has the unphysical feature of changing
sign for physically interesting values of $\epsilon$ and $n$.

As noted earlier, to this order in $\epsilon$, the phase boundary $T=T_c (t)$ in Fig~\ref{f1} is
determined by the condition $R=0$ and yields the values for $T_c$ given in (\ref{restc1}) and
(\ref{restc2}) (the order $\epsilon^{1/2\nu}$ corrections follow from assuming the scaling form
(\ref{resRa}) and the fact that $F_3$ has an expansion in integer powers of $\epsilon$). The
result for
$\chi$ and
$R$ in this subsection are not valid in the region
$|T - T_c (t) | \ll T_c (t)$, where, instead, we have to insert the results for $R$, $K$, and $U$
in Section~\ref{sec:cc} into the tricritical crossover functions of
Appendix~\ref{tricritical}. In the ordered phase we have a second low $T$ limit of the CQFT
(region III of Fig~\ref{f1}) where again the properties are dominated by scales set by the
$t<0$, $T=0$ ground state ($\rho_s$ for $n>1$, and $\Delta_-$ for $n=1$). A separate analysis with
a spontaneously broken symmetry is necessary here; it can be easily performed by our methods, but
we have not presented it in this paper.

Let us now turn to dynamic properties.
At this
order, the self energy has no momentum or frequency dependence; as a result 
the imaginary part of the susceptibility contains only delta functions at real
frequencies:
\begin{equation}
\mbox{Im} \chi (k, \omega) = \frac{\pi}{2\varepsilon (k)}
\bigl[ \delta(\omega - \varepsilon(k) ) - \delta (\omega + \varepsilon (k)) \bigr]
\label{imchi1}
\end{equation}
with $\varepsilon^2 (k) = k^2 + \xi^{-2}$. This is
clearly an artifact of the one-loop result, as the spectral density is required on
general grounds to be non-zero at all frequencies at any non-zero temperature. 
The two-loop computation of the imaginary part of the susceptibility in
Section~\ref{twoloop} will not suffer from this defect.

\subsubsection{Response to a field coupling to the conserved $O(n)$ charge}
\label{sec:conserve}

The $O(n)$ symmetric action ${\cal S}$ possesses a set of $n(n-1)/2$ conserved Noether
charges. In this subsection we will examine the susceptibility, $\chi_H$, associated with an
external field $H$ which couples to one these charges. This analysis is motivated primarily by
recent work~\cite{CSY} on the $d=2$, $O(3)$ sigma model of two-dimensional quantum
antiferromagnets, where this susceptibility is the response to an ordinary uniform
magnetic field. Here, we will complement the earlier $1/n$ expansion results~\cite{CSY} by the
$\epsilon$ expansion. It is also worth noting here that what we have denoted here as the
ordinary susceptibility $\chi (k, \omega )$ is the staggered susceptibility of the quantum
antiferromagnet.

Let us orient the field $H$ such that it causes a precession of $\phi$ in the $1-2$
plane. The time derivative term in (\ref{cals}) is then modified to 
\begin{equation}
\frac{1}{2} \left[ (\partial_{\tau}\phi_1 - i H \phi_2 )^2 + 
 (\partial_{\tau}\phi_2 + i H \phi_1 )^2 + \sum_{\alpha=3}^{n} (\partial_{\tau}
\phi_{\alpha})^2 \right]
\label{cons1}
\end{equation}
The susceptibility, $\chi_H$, is the second derivative of the free energy with
respect to variations in $H$. We can evaluate this using the method described in
Section~\ref{sec:method} and Appendix~\ref{Tgt0}; to first order in $u_0$ we obtain
\begin{eqnarray}
\chi_H = && 2 T \sum_{\epsilon_n \neq 0} \int \frac{d^d q}{(2 \pi)^d}
\frac{q^2 + t_0 - \epsilon_n^2}{(\epsilon_n^2 + q^2 + t_0 )^2}
+ 2 T \int \frac{d^d q}{(2 \pi)^d} \frac{1}{q^2 + R} \nonumber \\
&& - 2 u_0 \left( \frac{n+2}{6} \right) \left[
T  \int \frac{d^d q_1}{(2 \pi)^d} \left( \sum_{\epsilon_n \neq 0} \frac{1}{\epsilon_n^2
+ q_1^2 + t_0} + \frac{1}{q_1^2 + R} \right)
- \int \frac{d^{d+1} p}{(2 \pi)^{d+1}} \frac{1}{p^2} \right] \nonumber\\
&&~~~~~~~~~~~~~~~~~~~~~~~~\times
\left[ T \sum_{\Omega_n \neq 0} \int \frac{d^d q_2}{(2 \pi)^d}
\frac{q_2^2 + t_0 - 3 \Omega_n^2}{(\Omega_n^2 + q_2^2 + t_0)^3} \right] \nonumber \\
&& - 2 u_0 \left( \frac{n+2}{6} \right) \left[
T  \int \frac{d^d q_1}{(2 \pi)^d} \left(\frac{1}{q_1^2 + R} - \frac{1}{q_1^2} \right)
\right]
\left[ T  \int \frac{d^d q_2}{(2 \pi)^d}
\frac{1}{(q_2^2 + R)^2} \right]
\label{cons2}
\end{eqnarray}
Evaluating the frequency summations and the momentum integrals, expressing in terms of
the dimensionless coupling $g$ (Eqn (\ref{defg})), and expanding some of the terms to
the needed order in $\epsilon$, we obtain from (\ref{cons2})
\begin{eqnarray}
\frac{\chi_H}{T^{d-1}} = && Q_d \left( \frac{t_0}{T^2} \right) + 
\frac{2 \Gamma(1-d/2)}{(4 \pi)^{d/2}} \left( \frac{R}{T^2} \right)^{d/2 - 1} 
+ \frac{g}{2} \left(\frac{n+2}{6} \right) \nonumber \\
&&~~~~
+ g \left(\frac{n+2}{6} \right)
\left( \frac{\mu}{T} \right)^{\epsilon} Q_d^{\prime} \left( \frac{t_0}{T^2} \right)
\left[ F_3 \left( \frac{t_0}{T^2} \right) - 2 \pi \left( \frac{R}{T^2} \right)^{1/2}
- \frac{t_0}{T^2} \frac{1}{\epsilon} + \frac{t_0}{2 T^2} \ln \frac{t_0}{T^2} \right]
\label{cons3}
\end{eqnarray}
where the function $F_d (y)$ was defined in (\ref{deffd}) and
\begin{equation}
Q_d (y) \equiv \int \frac{d^d k}{(2 \pi)^d} \frac{1}{2 \sinh^2 (\sqrt{k^2 + y}/2) }
- \frac{2 \Gamma(1-d/2)}{(4 \pi)^{d/2}} y^{d/2-1}
\label{cons4}
\end{equation}
A number of important results now follow from (\ref{cons3}) and (\ref{cons4}); as the
analysis is quite similar to that in Section~\ref{sec:cc}, we will omit the details:
\newline
({\em i\/}) After expressing in terms of the renormalized $t$ by
$t_0 = t ( 1 + g (n+2)/(6 \epsilon))$ (Eqns (\ref{deft}) and (\ref{defzs})), we find that
the poles in $\epsilon$ cancel to order $g$.
\newline
({\em ii\/}) The resulting expression for $\chi_H$ is then analytic as a function
of $t$ at $t=0$. This follows from the previously established analyticity of $R$ at
$t=0$, and the fact that $Q_d (y)$ is analytic at $y=0$. The result (\ref{cons3}) can
therefore be used both for $t>0$ and $t<0$.
\newline
({\em iii\/}) For $t>0$, express $t$ in terms of the true energy gap $\Delta_+$ (using
(\ref{valdelp})), and evaluate (\ref{cons3}) at the fixed point coupling $g= g^{\ast}$
(Eqn (\ref{valgast})). All dependence on the renormalization scale $\mu$ disappears,
and $\chi_H$ then satisfies the scaling form~\cite{CSY,conserve}
\begin{equation}
\chi_H = T^{d-1} \Psi_H \left( \frac{\Delta_+}{T} \right),
\label{cons5}
\end{equation}
where $\Psi_H$ is a universal function, easily obtainable from (\ref{cons3}).
A similar result holds for $t<0$, where the renormalized energy scale is now the spin
stiffness $\rho_s$, related to $t$ by (\ref{valn0}) and (\ref{valrhos}).

We will be a little more explicit at the critical coupling $t=0$. 
First, we have
\begin{eqnarray}
Q_d (0) &=& \frac{4 (d-1) \Gamma(d-1) \zeta (d-1)}{(4 \pi)^{d/2} \Gamma(d/2)}
\nonumber \\ Q_d^{\prime} (0) &=& \frac{1}{4 \pi^2} + {\cal O} (\epsilon )
\label{cons6}
\end{eqnarray}
Using these results, (\ref{valfd}) and (\ref{resRtgt0}), we get from (\ref{cons3}): 
\begin{equation}
\frac{\chi_H}{T^{d-1}} = \frac{1}{3} - \sqrt{\epsilon}
\left( \frac{n+2}{6 (n+8)} \right)^{1/2}
 + \epsilon\left(\frac{2(n+2)}{3(n+8)}  + 0.310311256\ldots \right)  +
{\cal O} (\epsilon^{3/2})
\label{cons7}
\end{equation}
at $t=0$.
At the physical value for two-dimensional antiferromagnets, $n=3$, $\epsilon=1$,
the successive terms in (\ref{cons7}) oscillate in sign, and do not become smaller -
so a direct evaluation does not yield a useful numerical estimate.

\subsection{Two loop results}
\label{twoloop}

All computations in this subsection will be limited to the critical coupling $t=0$.

Two loop results for the values of the static quantities $R$, $\chi^{-1} (0,0)$
and $\partial \chi^{-1} (k, 0)/\partial k^2 |_{k=0}$ are presented in Appendix~\ref{Tgt0}.
Our main purpose in obtaining the results is that they provide an explicit
demonstration of the consistency of the method proposed in this paper: all ultraviolet and
infrared divergences cancel as required, and the results take the form of a systematic series in
powers of $\sqrt{\epsilon}$, along with a finite number of factors of $\log (\epsilon)$.

In this subsection we will limit our discussion to dynamic observables, in particular those
related to $\mbox{Im} \chi (k, \omega)$. Two loop contributions make an important qualitative
difference in that the delta functions peaks in (\ref{imchi1}) are broadened due to
dissipative thermal effects. 

We begin with the expression (\ref{reschi}), retaining only the terms dependent upon
the external frequency, setting the coupling $u_0$ to its fixed point value, and keeping terms
up to formal order $\epsilon^2$:
\begin{eqnarray}
\chi^{-1} (k, i\omega_n) = && Z \omega_n^2
-\epsilon^2 \frac{2(n+2)}{(n+8)^2}  \frac{T^2}{S_4^2}\sum_{\epsilon_n,\Omega_n} 
\int
\frac{d^3 q_1}{(2\pi)^3} \frac{d^3 q_2}{(2\pi)^3} 
 \frac{1}{(q_1^2 + \widetilde{\sigma}(\epsilon_n))(q_2^2 +
\widetilde{\sigma}(\Omega_n))} \nonumber \\
&&~~~~~~~~~~\times \frac{1}{((k-q_1 -q_2)^2 +
\widetilde{\sigma}(\omega_n - \epsilon_n - \Omega_n))}
+\mbox{~~terms independent of $\omega_n$,}
\label{imchi2}
\end{eqnarray}
where at the critical coupling $t=0$ (compare (\ref{restsigma}))
\begin{equation}
\widetilde{\sigma} (\epsilon_n) = \epsilon_n^2 + R \delta_{\epsilon_n, 0}
\label{imchi3}
\end{equation}
Notice that the finite frequency propagators in (\ref{imchi2}) have only their bare mass which
vanishes at the critical coupling $t=0$, while the zero frequency propagator has a
fluctuation-induced mass of order $\epsilon T^2$. Clearly, this distinction is an artifact of
our method which treats the zero frequency modes in a manner distinct from the finite frequency
modes. However, the distinction is unimportant in an expansion for physical quantities as a
series in $\sqrt{\epsilon}$ (modulo logarithms of $\epsilon$), for the finite frequency modes
have a minimum value of
$\omega_n^2 = 4 \pi T^2$, which overwhelms any mass term of order $\epsilon T^2$ we might
consider adding to their propagator. In other words, (\ref{imchi2}) provides the leading
frequency-dependent contribution to $\chi^{-1} (\omega_n)$ in an expansion in $\sqrt{\epsilon}$
for all physically allowed values of $\omega_n$. 

We are interested here in the value of $\mbox{Im} \chi (k, \omega)$ for real frequencies
$\omega$. In principle, this can be obtained by analytic continuation from the values of the
susceptibility at the Matsubara frequencies. However, and this is a key point, there is no
guarantee that the analytically continued result will also be a systematic series in
$\sqrt{\epsilon}$, valid for all values of $\omega$. In fact, it is not difficult to see that
the analytically continued result is valid only for $\omega \gg \sqrt{\epsilon} T$.
This condition can be traced to the ambiguity in the mass term for the finite frequency
propagators discussed in the previous paragraph; while this ambiguity is unimportant at the
Matsubara frequencies, a simple estimate shows that it strongly modifies $\mbox{Im} \chi (k,
\omega)$ for $\omega \sim \sqrt{\epsilon} T$.
As a result, we are only able to obtain here systematic results for $\mbox{Im} \chi (k, \omega)$
for $\omega \gg \sqrt{\epsilon} T$. The $\epsilon$ dependence of the very important low frequency
limit $\omega \rightarrow 0$ for finite $T$ remains an open problem. Similar difficulties were
also encountered earlier in the $1/N$ expansion~\cite{CSY} of the same problem, where
the expansion broke down for $\omega \sim T$. Here, we are able to explore the
region  $\sqrt{\epsilon} T \ll \omega \ll T$, and in particular have systematic results
for $\omega \sim T$. 

In the remainder of the section we will therefore restrict our attention to $\omega \gg
\sqrt{\epsilon} T$. Under these conditions we can drop the mass $R$ even from the zero
frequency propagators while computing the imaginary part (all infrared divergences controlled
by a finite $R$ occur only in the real part). Evaluating the frequency summation in 
(\ref{imchi2}), analytically continuing to real frequencies, and taking the imaginary part,
we obtain at $k=0$
\begin{eqnarray}
&& \mbox{Im} \chi^{-1} (k=0, \omega > 0 ) =  -\epsilon^2 \frac{\pi (n+2)}{(n+8)^2} 
\frac{1}{S_4^2} \int \frac{d^3 q_1}{(2\pi)^3} \frac{d^3 q_2}{(2\pi)^3} 
\Biggl[ \nonumber \\
&&~~~~~~\frac{ 3 [ n (|q_1 + q_2|) ( 1 + n(q_1) + n(q_2) ) - n(q_1) n(q_2) ]}{
4 |q_1 + q_2| q_1 q_2} \delta(\omega + |q_1 + q_2| - q_1 - q_2 ) \nonumber \\
&&~~~~~~\left.
+ \frac{1 + (3/2) (n(q_1) + n(q_2))(1 +  n (|q_1 + q_2))}{ 4 |q_1 + q_2| q_1 q_2} \delta(\omega
- |q_1 + q_2| - q_1 - q_2 ) \right]~~~~\omega \gg \sqrt{\epsilon} T
\label{imchi4}
\end{eqnarray}
with $\mbox{Im} \chi^{-1} (k, \omega < 0 ) = - \mbox{Im} \chi^{-1} (k, - \omega )$,
and where $n(x) = 1/(e^{x/T} - 1)$ is the Bose function.
A similar result can also be obtained for $k \neq 0$ but we will refrain from displaying it; we
will limit ourselves to analyzing the simpler $k=0$ result. The angular integrals in
(\ref{imchi4}) can be performed and we obtain then
\begin{eqnarray}
 && \mbox{Im} \chi^{-1} (0, \omega ) =  -\epsilon^2 \frac{2 \pi (n+2)}{(n+8)^2} 
\left[ 3 \int_{\omega/2}^{\infty} d q_1 \int_{\omega_2}^{\infty} d q_2
\left[ n (q_1 + q_2 - \omega) ( 1 + n(q_1) + n(q_2) ) - n(q_1) n(q_2) \right] \right. \nonumber
\\ 
&&~~~~~~~~~~~~~~~~\left. + \int_{0}^{\omega/2} d q_1 \int_{\omega_2 - q_1}^{\omega_2} d q_2
\left[1 + \frac{3}{2}(n(q_1) + n(q_2))(1 + n (q_1 + q_2 - \omega))
\right] \right]~~~~\omega \gg \sqrt{\epsilon} T
\label{imchi5}
\end{eqnarray}
Somewhat unexpectedly, all of the integrals in (\ref{imchi5}) can also be performed
analytically; after a lengthy, but straightforward, computation we obtained a final result which
had a surprisingly simple form---we got
\begin{equation}
\mbox{Im} \chi^{-1} (0, \omega ) =  -\epsilon^2 \frac{2 \pi (n+2)}{(n+8)^2}
\left[ \frac{\omega^2}{8} + \pi^2 T^2 + 6 T^2 \mbox{Li}_2 ( e^{-\omega/2T}) \right]
~~~~\omega \gg \sqrt{\epsilon} T
\label{imchi6}
\end{equation}
where $\mbox{Li}_2 (x) $ is the dilogarithm function 
\begin{equation}
\mbox{Li}_2 (x) = -\int_0^{x} \frac{dy}{y} \ln (1-y).
\label{defdilog}
\end{equation}
For large $\omega$ we have from (\ref{imchi6})
\begin{equation}
\mbox{Im} \chi^{-1} (0, |\omega| \rightarrow \infty )
= \frac{\pi \eta \omega^2}{2} \mbox{sgn} (\omega)
\label{imchi7}
\end{equation}
where $\eta = \epsilon^2 (n+2)/(2 (n+8)^2)$ is the field anomalous
dimension; this is precisely the result expected~\cite{CSY} at this order from scaling.
For small $\omega$, the result (\ref{imchi6}) taken
at face value gives us
\begin{equation}
\mbox{Im} \chi^{-1} (0, \omega \rightarrow 0) = -\epsilon^2 \frac{2 \pi (n+2)}{(n+8)^2}
\left[ 2 \pi^2 T^2 \mbox{sgn} (\omega) + 3 \omega T\left(\ln(|\omega|/2T) - 1 \right) \right]
\label{imchi8}
\end{equation}
This singular behavior at small $\omega$ is clearly an artifact of taking (\ref{imchi6}) beyond
its regime of validity; we expect instead that $\mbox{Im} \chi^{-1} (0, \omega)
\sim \omega$ for small $\omega$, but have no direct method here for estimating its coefficient.

One measure of the strength of the dissipation computed above is the value of
$\mbox{Im} \chi^{-1} (0, \omega = T)$, where our expansion is expected to be reliable.
This is characterized by the damping rate $\Gamma_{RT}$ defined in (\ref{defgammat}).
From (\ref{imchi6}) and (\ref{reschioneloop}) we have to leading order in $\epsilon$
\begin{eqnarray}
\frac{T}{\Gamma_{RT}} &=& \epsilon \frac{3}{(n+8) \pi}
\left[ \frac{1}{8} + \pi^2  + 6  \mbox{Li}_2 ( e^{-1/2}) \right] \nonumber \\
&=& 13.770249 \frac{\epsilon}{n+8}
\label{imchi10}
\end{eqnarray}
It is interesting to compare this value with the exact result for the one-dimensional
transverse-field Ising ($n=1$, $\epsilon = 2$), for which we get~\cite{statphys}
\begin{eqnarray}
\frac{T}{\Gamma_{RT}} =&& \frac{1}{2 \pi^2} 
\left( \frac{\Gamma(1/16)}{\Gamma(15/16)} \right)^2
 \left| \Gamma \left(\frac{15}{16} - \frac{i}{4 \pi} \right) \right|^4
\sin ( \pi/8) \sinh(1/2) 
=  2.560527\ldots \nonumber \\
&&~~~~~~~~~~~~~~~~~~~~~~~~~~~~~~~~~~~~~\mbox{exact value for $n=1$, $\epsilon=2$}\nonumber
\\
\frac{T}{\Gamma_{RT}} =&& 3.06\ldots~~~~\mbox{$\epsilon$ expansion (\ref{imchi10}) at $n=1$,
$\epsilon=2$}
\label{imchi11}
\end{eqnarray}
(the $\Gamma$ functions on the right hand side, should not be confused with the damping rate
$\Gamma_{RT}$). The agreement is quite reasonable, even for $\epsilon$ as large as 2.

It is interesting to compare the ratio of relaxation rates at $\omega = T$ ($\Gamma_{R}$,
defined in (\ref{defgammar})) with
that at $\omega = 0$ ($\Gamma_{RT}$) for the $n=1$, $d=1$ case, where we have results for both.
We obtain~\cite{statphys}
\begin{eqnarray}
\frac{\Gamma_{RT}}{\Gamma_{R}} &=& \frac{\pi^2}{2 \sin^2 (\pi /16) \sinh(1/2) } \left(
\frac{\Gamma(15/16)}{\Gamma(1/16)} \right)^2
\left| \Gamma \left(\frac{15}{16} - \frac{i}{4 \pi} \right) \right|^{-4} \nonumber \\
&=& 
0.981700183338266\ldots  
\end{eqnarray}
Notice that the two rates are almost exactly equal, as had been conjectured earlier for $d=2$,
$n=3$ in Ref~\cite{CSY}. We suspect that the near equality is quite general, and so
$\Gamma_{RT}$ is always a good estimate for $\Gamma_T$ in the high $T$ limit (region II of
Fig~\ref{f1}).

\section{Models above their upper critical dimension}
\label{sec:above}

The computations now follow the same basic strategy as that used for systems below their upper
critical dimension in Section~\ref{sec:below}. The main difference is that the expansion is now
in terms of the bare value of the irrelevant non-linearity $u_0$, rather than its universal
fixed-point value. Further, there are no non-trivial renormalizations, and the renormalization
constants $Z$, $Z_1$, $Z_2$, $Z_4$ can all be set equal to unity. The results now will have
some explicit, non-universal, cutoff dependence which cannot be removed by a simple
renormalization: this is because the
$T=0$ quantum critical point is above its upper-critical dimension, and the field theory is
therefore non-renormalizable.

We will analyze a class of
models with the general effective action, ${\cal S}_g$, of the form
\begin{equation}
{\cal S}_g = \frac{T}{2} \sum_{\omega_n} \int \frac{d^d k}{( 2 \pi)^d}
|\phi(k, \omega_n)|^2 \left[ M(\omega_n) + k^2 + t_0 \right]
+ \frac{u_0}{4!} \int_0^{1/T} d \tau \int d^d x \phi^4 ( x, \tau),
\label{up1}
\end{equation}
where we have the usual Fourier-transformed field
\begin{equation}
\phi(k, \omega_n) = \int_0^{1/T} d\tau \int d^d x \phi(x, \tau) e^{i(kx - \omega_n \tau)}.
\label{up2}
\end{equation}
Different choices for $M(q, \omega_n)$ describe a variety of physical situations:\newline
(a) $M(\omega_n) = \omega_n^2$: This obviously corresponds to the action ${\cal S}$ of a
quantum rotor ($n \geq 2$) or transverse-field Ising ($n=1$) model which we have already
studied in Section~\ref{sec:below}. It has dynamic exponent $z=1$ and upper-critical
dimension $d=3$.\newline 
(b) $M(\omega_n) = -i D \omega_n$: Now ${\cal S}_g$ describes a dilute bose
gas~\cite{otherbose,cres,weichmann,dsf}, with
$D$ a constant (analogous to the velocity $c$ for ${\cal S}$) related to the mass of the bosons.
The dynamic critical exponent is $z=2$, and the upper-critical dimension is $d=2$.\newline
(c) $M(\omega_n) = D |\omega_n|$: In this case, ${\cal S}_g$ describes spin fluctuations
in the vicinity of the onset of spin-density wave order in a Fermi liquid~\cite{hertz}. Unlike
the cases I and II, the $T=0$ dynamic susceptibility now does not have a quasiparticle pole in
the paramagnetic phases, but instead has a cut describing the particle-hole continuum.
The dynamic critical exponent is $z=2$ and the upper-critical dimension is $d=2$. Finally, it
must be noted that in this case ${\cal S}_g$ is only applicable in the paramagnetic (or Fermi
liquid) phase~\cite{scs}; a separate action is needed within the magnetically ordered
phase.\newline 
Note that all of the above choices for $M$ share the property $M(0) = 0$. Also in
all three cases the correlation length exponent $\nu = 1/2$, and the coupling
$u_0$ is irrelevant at the $u_0 = 0$ quantum critical point with scaling dimension $-\theta_u$;
for the models above we have $\theta_u = d+z-4$, a relationship which is not always valid.

Another model of interest is the quantum-critical point describing the onset of ferromagnetism
in a Fermi liquid~\cite{hertz}. It has recently been pointed out~\cite{belitz} the effective
action now contains non-analytic dependencies on the momentum, $k$, ($\sim k^{d-1}$ in clean
systems and
$\sim k^{d-2}$ in random systems) which are present only at
$T=0$. This singular behavior is possible because gapless fermion modes are being integrated
out. It is now clearly necessary to also account for the $T$ dependence arising from the
elimination of the critical fermion models. This should be possible using the general methods of
this paper, but this issue shall not be addressed in this paper.

Returning to models (a)-(c) above, we perform a perturbation theory in $u$ as described in
Sections~\ref{sec:method} and~\ref{sec:below}. The generalization of (\ref{resR}) to linear
order in $u_0$ is now
\begin{equation}
R = t_0 + u_0 \left( \frac{n+2}{6} \right)
\int \frac{d^d q}{(2 \pi)^d} \left( T \sum_{\epsilon_n \neq 0} \frac{1}{M(\epsilon_n) + q^2 +
t_0} + \frac{T}{q^2}  - \int \frac{d\epsilon}{2 \pi} \frac{1}{M(\epsilon) + q^2} \right).
\label{up3}
\end{equation}
The susceptibility, defined in (\ref{chidef}) is obtained from $R$ by generalizing
(\ref{reschioneloop}) to
\begin{equation}
\chi^{-1} (k, i \omega_n) = k^2 + M( \omega_n ) + R - u_0 T \left(\frac{n+2}{6} \right)
\frac{2 \Gamma((4-d)/2)}{(d-2)(4 \pi)^{d/2}} R^{(d-2)/2}
\label{up4}
\end{equation}
We have assumed above, and will continue to assume below that $2<d<4$. The correlation length,
$\xi$, is given at this order by $\xi^2 = \chi (0,0)$, as in Section~\ref{sec:susc}. To apply the
analog of (\ref{identity}), it is convenient to separate
$R$ into the following form:
\begin{equation}
R = t_0 + u_0 \left( \frac{n+2}{6} \right) \left[ R_1 + R_2 + R_3 \right]
\label{up5}
\end{equation}
with
\begin{eqnarray}
R_1 &=& \int \frac{d^d q}{(2 \pi)^d} \left(T \sum_{\epsilon_n} 
\frac{1}{M(\epsilon_n) + q^2 + t_0}  - \int \frac{d\epsilon}{2 \pi}
\frac{1}{M(\epsilon) + q^2 + t_0} \right) \nonumber \\
R_2 &=& - T\int \frac{d^d q}{(2 \pi)^d} \left( \frac{1}{q^2 + t_0} - \frac{1}{q^2} \right)
\nonumber \\
R_3 &=& \int \frac{d^d q}{(2 \pi)^d} \int \frac{d\epsilon}{2 \pi}\left(
\frac{1}{M(\epsilon) + q^2 + t_0} - \frac{1}{M(\epsilon) +
q^2}
\right)
\label{up6}
\end{eqnarray}
The integral in $R_1$ is ultraviolet convergent in all $d$ in models (a) and (b),
and is convergent for $d<4$ in model (c). The integral in $R_2$ is ultraviolet convergent
for $d<4$. All of the ultraviolet divergences have been isolated in $R_3$. 
For all models (a)-(c) this divergence can be separated by a single subtraction which is a
linear function $t_0$; we write $R_3$ as
\begin{eqnarray}
R_3 = && \int \frac{d^d q}{(2 \pi)^d} \int \frac{d\epsilon}{2 \pi}\left(
\frac{1}{M(\epsilon) + q^2 + t_0} - \frac{1}{M(\epsilon) +
q^2} + \frac{t_0}{(M(\epsilon) + q^2)^2}\right) \nonumber \\
&&~~~~~~~~~~~~~~~~~- \int_0^{\Lambda} \frac{d^d q}{(2
\pi)^d}
\int
\frac{d\epsilon}{2 \pi}\frac{t_0}{(M(\epsilon) + q^2)^2}
\label{up7}
\end{eqnarray}
The last integral is a cutoff ($\Lambda$) dependent term which has the simplifying feature of being
a  linear (and therefore analytic) function of $t_0$. The first integral is now ultraviolet
convergent, but is a more complicated function of $t_0$. In
other models additional subtractions involving higher powers of $t_0$ may be necessary at this
stage. 

\begin{figure}
\epsfxsize=14in
\centerline{\epsffile{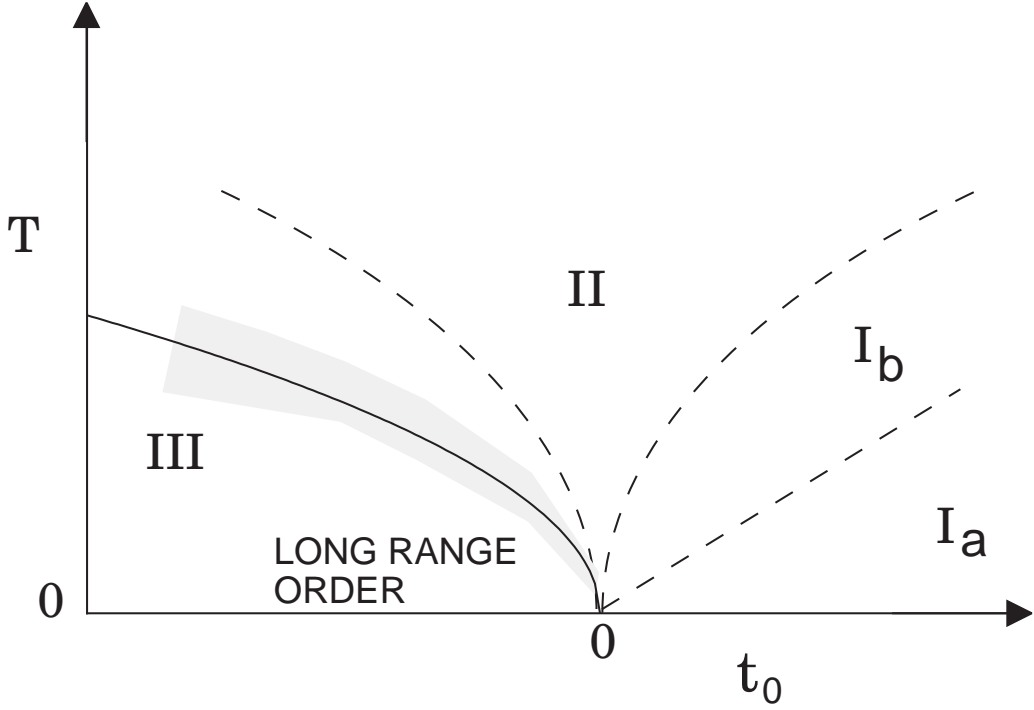}}
\vspace{0.5in}
\caption{
As in Fig~\protect\ref{f1}, but for the case where the quantum critical point is above its
upper-critical dimension. Now properties are controlled by the value of the coupling $u_0$,
associated with the least irrelevant operator involving quantum-mechanical interactions; we denote
its scaling dimension as $-\theta_u$ ($\theta_u > 0$). The crossover boundaries on either side of
region II, and the phase transition line $T=T_c (t)$, now scale  as
$T \sim (|t|/u_0)^{z
\nu/(1 +\theta_u \nu)}$. 
The boundary between regions Ia and Ib scales as $T \sim t^{z\nu}$. The expansion is now in powers
of the bare coupling $u_0$, and as before, classical tricritical crossover functions are needed in
the shaded region 
$|T - T_c (t)|
\ll T_c (t)$. For more discussion on the regions see the discussion below Eqn
(\protect\ref{up7a}). }
\label{f3}
\end{figure}

Carrying out all the frequency integrals and summations in (\ref{up6}) and (\ref{up7}) for
(along with the momentum integration of the last term in (\ref{up7}), we find that in all three
models
$R$ takes the form
\begin{equation}
R = t_0 ( 1 - u_0 c_1 \Lambda^{\theta_u} ) + u_0 c_2 T^{(1 + \theta_u \nu)/z \nu}
\Upsilon \left( c_3 \frac{t_0}{T^{1/z \nu}} \right)
\label{up7a}
\end{equation}
where $c_1$, $c_2$, and $c_3$ are constants, and $\Upsilon$ is a {\em universal\/} scaling
function given by $R_1 + R_2 + R_3$ but with the last term in (\ref{up7}) omitted.
By examining the
limiting behavior of (\ref{up4}) and (\ref{up7a}) we can delineate the different physical regimes
as shown in Fig~\ref{f3}~\cite{weichmann,millis,oppermann}.
\newline
(I) $T \ll (t_0/u_0)^{z \nu/(1 + \theta_u \nu)}$: low $T$ limit of CQFT; paramagnetic phase
\newline
In this regime we are dominated by the $t>0$ ground state, with given $\xi$ given to leading
order by its $T=0$ value. The subleading temperature dependent corrections are however different
depending upon whether the argument of the scaling function $\Upsilon (y)$ is small or large.
These sub-regimes are therefore:\newline
(Ia) $T \ll t_0^{z\nu}$\newline
The nature of the corrections depends upon the behavior of $\Upsilon (y \rightarrow \infty)$,
which can vary considerably from model to model. In models (a) and (b) the $t_0 > 0$ ground state
has a gap, so the leading correction will be exponentially small in temperature. Model (c) is a
Fermi liquid for $t_0 > 0$, and has power-law corrections in $T$ which will be described in more
detail below\newline
(1b) $t_0^{z\nu} \ll  T \ll t_0^{z\nu/(1 +\theta_u\nu)}$
\newline
Now the $T$-dependent corrections involve $\Upsilon (0)$, which is always a pure number; so we
have
\begin{equation}
\xi^{-2} (T) = \xi^{-2} (T=0) + u_0 c_2 T^{(1 + \theta_u \nu)/z \nu}
\Upsilon (0) + \ldots
\label{up7b}
\end{equation}
(II) $ T \gg |t_0|^{z\nu/(1 + \theta_u \nu)}$: high $T$ limit of CQFT
\newline
Now $T$ is the most important energy scale and all $t_0$-dependent corrections can be neglected.
The correlation length still obeys (\ref{up7b}) but with the second term now being larger.

The transition to the ordered state occurs, as before, at $R=0$. To leading order in $u_0$,
$T_c$ is given by (\ref{up7a}) to be $T_c = (|t_0| /(u_0 c_2 \Upsilon (0))^{z\nu/(1 + \theta_u
\nu)}$.

In the remaining presentation  we specialize to model (c); the properties of models
(a)-(b) will be quite similar. The upper critical dimension is $d=2$, and we assume we are above
it. For this case, the explicit result for $R$ is 
\begin{equation}
R = t_0 \left( 1 -  \frac{u_0 \Lambda^{d-2}}{D} \left( \frac{n+2}{6} \right)
\frac{S_d}{\pi (d-2)} \right)
+ \frac{u_0}{D} (TD)^{d/2} 
\left( \frac{n+2}{6} \right) \Upsilon \left( \frac{t_0}{DT} \right),
\label{up8}  
\end{equation} 
where $\Upsilon$ is initially obtained as
\begin{equation}
\Upsilon(y)=
\frac{1}{\pi} \int \frac{d^d q}{(2 \pi)^d} \left[
\ln\left( \frac{q^2}{2 \pi} \right) - \frac{2 \pi}{q^2 + y} - \psi\left( \frac{q^2 + y}{2 \pi}
\right) + \frac{\pi + y}{q^2} \right]
\label{up9}
\end{equation}
where $\psi$ is the digamma function; it can be verified that the integral over $q$ in
(\ref{up9}) is convergent for $2<d<4$. Now we use the identity
$\psi(z+1) =
\psi(z) + 1/z$ to simplify (\ref{up9}) to
\begin{equation}
\Upsilon(y)=
\frac{1}{\pi} \int \frac{d^d q}{(2 \pi)^d} \left[
\ln\left( \frac{q^2}{2 \pi} \right) - \psi\left(1+ \frac{q^2 + y}{2 \pi}
\right) + \frac{\pi + y}{q^2} \right].
\label{up10}
\end{equation}
In this form, it is manifestly clear that $\Upsilon (y)$ is analytic as a function of $y$ at
$y=0$, and so from (\ref{up8}), $R$ is analytic at $t_0 = 0$. Indeed the first
singularity of (\ref{up10}) is at $y=-2\pi$, and (\ref{up10}) can be used for all $y > - 2
\pi$. This allows us to access the region with $t_0 < 0$, but $T > T_c (t_0)$. The singularity
at $y= -2\pi$ is of no physical consequence, as it is well within the ordered phase.
Recall that a similar phenomenon occurred in our earlier analysis of ${\cal S}$ for $d<3$ in
Section~\ref{sec:cc}. 

By evaluating the large $y$ behavior of $\Upsilon (y)$, we can determine from (\ref{up4}) the
$T$-dependent corrections in regime 1a. We find
\begin{equation}
\xi^{-2} (T) = \xi^{-2} (T=0) + \frac{u_0 D}{t_0^{(4-d)/2}} \frac{(n+2) \Gamma((4-d)/2)}{36 (4
\pi)^{d/2}} T^2
\end{equation}
As noted earlier, the $T$-dependent correction is a power-law characteristic of a Fermi liquid.

\section{Conclusions}
\label{conc}

This paper has provided a general strategy computation of finite temperature 
universal crossover functions near quantum-critical points. The strategy can be broken down
into steps, each step containing distinct physical effects; this separation is an important
advantage of our method. The steps are:\newline
({\em i\/}) Renormalize the $T=0$ CQFT to obtain a well-defined quantum theory whose ground
state, excited states, and scattering amplitudes between them, are known. In principle, this
information completely specifies the non-zero $T$ properties, and no further
renormalizations should be necessary.\newline 
({\em ii\/}) Use the information in ({\em i\/}) to integrate out all degrees of freedom with a
finite Matsubara frequency, to derive effective action (which could be quite complicated) for
the zero frequency mode.\newline
({\em iii\/}) Analyze the effective action by an appropriate technique of classical statistical
mechanics.\newline
We have applied this method in this paper to a relativistic $n$-component $\phi^4$ theory below
its upper-critical dimension, and to a class of models above their upper-critical dimension.

We will now comment on the relationship of our results to some earlier work.

The early work on quantum-critical points~\cite{hertz,suzuki,cres} studied only the
quantum-to-classical crossover in the shaded region of Figs~\ref{f1} and~\ref{f3}.
The crossovers in the remainder of the phase diagram, and their universal properties, were
missed.

Rasolt {\em et. al.\/} studied the quantum-to-classical crossovers in the dilute Bose gas in
$d=3$. In the present language, these are the crossovers near the $T=0$
quantum-critical point at chemical potential $\mu=0$. They described the physics in terms of
the Gaussian-Heisenberg crossover of $\phi^4$ field theory in $d$-dimensions. This is closely
related, but not identical, to our description in terms of tricritical functions, as the latter
are the universal limit of the former when $u_0 \ll \Lambda^{4-d}$~\cite{nr}, where $\Lambda$ is
a momentum cut-off. Our approach properly identifies all the non-universal cut-off dependence as
due to the
$T=0$ quantum theory (in $R_3$ in Eqn (\ref{up6})), and shows that the finite temperature
corrections are universal (the scaling function $\Upsilon$ in (\ref{up7a})). Below the
upper-critical dimension, there are no non-universal cut-off dependencies, and the use of
tricritical crossovers is essential.

Another popular approach to the study of finite temperature crossovers has been the
momentum-shell renormalization group (RG), in which the RG equations are $T$-dependent and $T$
is itself scale dependent~\cite{hertz,dsf,CHN,millis,sss}. This method has been quite useful
in identifying qualitative features of the crossovers in static and thermodynamic quantities. 
However, quantitative crossover functions have been quite difficult to obtain. We
believe this is not merely a technical difficulty, but an intrinsic problem with the physical
basis of this approach. The dynamic consequences of quantum and thermal fluctuations are
physically quite distinct, and it appears quite 
unsound to interpolate between them by defining a
scale-dependent temperature. It is quite clear that such a method will not correctly
describe the thermal dissipative dynamics. Instead, our point of view is that the RG
flows are more properly considered as properties of the $T=0$ theory, and allow one to define
its eigenstates and $S$ matrices. The finite $T$ physics is then completely determined by
these properties. 

O'Connor and Stephens~\cite{oconnor} have used an idea similar to that in the
momentum-shell RG, but in the framework of the field-theoretic RG. They achieve this by 
defining some unusual renormalization conditions which seem designed to yield $\beta$ functions
which are temperature dependent. The physical meaning or mathematical justification of these
renormalization conditions is not clear to us, and our critiques in  the previous paragraph
apply here too. We also note that their results are not systematic expansion in some control
parameter (many of the terms contain $\epsilon$ to all orders), and are not naturally expressed
in the terms of renormalized $T=0$ energy scales which expose the full universality of
the physics.

Large $n$ expansions have been also been used to study finite temperature properties of quantum
critical points~\cite{CSY,bray}. 
They have the advantage of being uniformly valid over the entire
phase diagram. Their most extensive application has been in $d=2$~\cite{CSY},  where $T_c (t) =
0$ for large $n$. However $T_c$ is non-zero for $d>2$, and a large $n$ computation then gives
results consistent with those of this paper.

We now turn to discussing some open problems and directions for future research.

The major gap in existing results is a quantitative and systematic theory for the low-frequency
dynamics. This is an experimentally important question, as the damping rate directly determines
NMR relaxation rates in two-dimensional quantum antiferromagnets~\cite{CSY}
Our present approach fails for $\omega < \sqrt{\epsilon} T$, but it is possible that
a systematic analysis of a self-consistent approach, with $\epsilon$ a control parameter, 
can be performed.

The present paper has avoided discussion of logarithmic corrections in special
dimensions, either due to $d$ being the upper-critical dimension of the quantum-critical point,
or because of the logarithmic corrections that appear in weakly-coupled classical theories in
$d=2$. We made this choice to streamline our discussion, but it should not be too difficult to
extend our results to include these cases.

Finally, we have already noted that there should be interesting finite $T$ crossovers in nearly
ferromagnetic Fermi liquids, as some novel non-analytic $q$ dependencies in the effective action
have recently been pointed out~\cite{belitz}.

\acknowledgements
I thank E. Brezin, A.V. Chubukov, K. Damle and T. Senthil for helpful discussions, and A.V. Chubukov for valuable remarks on the manuscript.
J.~Ye collaborated with the author at the very early stages of this work.
This research was supported by the National Science Foundation 
under Grants DMR-96-23181 (at Yale) and 
PHY94-07194 (at the Institute for Theoretical Physics, Santa Barbara). 

\appendix

\section{Classical tricritical crossover functions}
\label{tricritical}
In this section we will tabulate results on the classical tricritical crossover functions
needed in the region $|T - T_c (t) | \ll T_c (t)$. 
We will confine our attention to the crossover function, $\Psi_{C} (q, v)$, 
appearing in (\ref{psicl}), for the static susceptibility. In the weak 
coupling region, $v \ll 1$, we can easily expand in a power series in 
integer powers of $v$:
\begin{equation}
\Psi_{C}^{-1} (q, v) = q^2 + 1 - \left( \frac{n+2}{6} \right) \frac{2 
\Gamma((4-d)/2)}{(d-2) (4 \pi)^{d/2}} v + {\cal O} ( v^{2} )
\label{tri0}
\end{equation}
In this region the tricritical crossovers connect smoothly with our 
$\epsilon$ expansion results in the region outside $| T - T_c (t) | 
\ll T_{c}(t)$; the equivalent of (\ref{tri0}) was already used in 
(\ref{reschioneloop}) and (\ref{up4}).

Within $| T - T_c (t) | 
\ll T_{c}(t)$, $v$ becomes large, and alternative perturbative expansions 
are needed to obtain tricritical crossovers. We will discuss two such 
methods here.

The first method is the expansion in $4-d$. The reader may be bothered by our simultaneous use
of an expansion in $\epsilon \equiv 3-d$ in the analysis of ${\cal S}$ in the main part of the
paper. However, the two expansions occur in separate calculations and compute
entirely different crossover functions. They are combined only in the final result, in which
the results of one appear as arguments of the other. So there is no inconsistency, and the
procedure is entirely systematic. The result for $\Psi_C$ can be read off
from earlier results~\cite{nr}; to leading order in $4-d$ we have
\begin{equation}
\Psi_C^{-1} (q, v) = q^2 + \left[ 1 + \frac{(n+8) v}{ 48 \pi^2 (4-d)} \right]^{-(n+2)/(n+8)}.
\label{tri1}
\end{equation}
When combined with (\ref{psicl}), we see that the critical point is at $R=0$; this will change
at higher orders in $4-d$, when we expect a critical value 
$R \sim [TU/K^{d/2}]^{2/(4-d)}$.

The function $\Psi_C$ can also be obtained in a large $n$ expansion, with $d$ now arbitrary.
Taking the large
$n$ limit of (\ref{scl}) while keeping $n v$ fixed, a straightforward calculation gives to
leading order
\begin{equation}
\Psi_C^{-1} (q, v) = q^2 + \Pi (v),
\label{tri2}
\end{equation}
where $\Pi (v)$ is determined by the solution of the non-linear equation
\begin{equation}
\Pi (v) + n v \frac{\Gamma((4-d)/2)}{3 (d-2) (4 \pi)^{d/2}} \left[\Pi (v)\right]^{(d-2)/2} = 1.
\label{tri3}
\end{equation}

Finally, it is easily checked that (\ref{tri0}), (\ref{tri1}) and (\ref{tri2}) 
all agree with each other in their mutually overlapping regimes of 
validity.

\section{Computations for ${\cal S}$ at $T=0$}
\label{zeroT}

We consider properties of the  model ${\cal S}$ (Eqn (\ref{cals})) at $T=0$
and $d<3$, in an expansion in powers of $\epsilon = 3-d$. 
We will compute the renormalized $T=0$ parameters which characterize the ground
state, and appear as arguments of the quantum-critical scaling functions. The
computations are standard~\cite{bgz,zj}, and we will be quite brief.

The renormalization constants $Z$, $Z_4$ (and $Z_2$) to the needed order in $g$ in the 
minimal subtraction scheme are:
\begin{eqnarray}
Z &=& 1 - \frac{n+2}{144 \epsilon} g^2 \nonumber \\
Z_2 &=& 1 + \frac{n+2}{6 \epsilon} g \nonumber \\
Z_4 &=& 1 + \frac{n+8}{6 \epsilon} g
\label{defzs}
\end{eqnarray}
The fixed point on the $\beta$-function is at $g = g^{\ast}$ with
\begin{equation}
g^{\ast} = \frac{6 \epsilon}{n+8} \left[
1 + \epsilon \frac{20 + 2 n - n^2}{2(n+8)^2}  \right].
\label{valgast}
\end{equation}

We consider the cases $t>0$ and $t<0$ separately.

\subsection{$t>0$}

At $T=0$, all properties are ``relativistically'' invariant, and are most
conveniently expressed in terms of a Euclidean momentum $p \equiv (\omega, k)$.
The renormalized susceptibility $\chi$ takes the form
\begin{equation}
\chi^{-1} (p)  = p^2 + t - \Sigma (p^2)
\end{equation}
where $\Sigma$ is the self energy.
The quasi-particle pole occurs at $p^2 = - \Delta_+^2$ which is the solution of
$\Delta_+^2 = t - \Sigma(-\Delta_+^2 ) $. The residue at this pole, ${\cal A}$ is
given by 
\begin{equation}
{\cal A} = \left( 1 - \left. \frac{\partial \Sigma}{\partial p^2} \right|_{p^2 = -
\Delta_+^2} \right)^{-1}.
\end{equation}
To leading order in $g$, we can now easily obtain by the usual methods
\begin{equation}
\Delta_+^2 = t \left( 1 + \frac{n+2}{6 \epsilon} g \right)
+ \frac{n+2}{6} \frac{\mu^{\epsilon}}{S_{d+1}} g \int \frac{d^{d+1}p}{(2 \pi)^{d+1}}
\left( \frac{1}{p^2 + t} - \frac{1}{p^2} \right)
\end{equation}
Evaluating this at $g=g^{\ast}$ we obtain
\begin{equation}
\Delta_+^2 = \mu^2 (t/\mu^2)^{2\nu}
\label{valdelp}
\end{equation}
where $\nu = 1/2 + \epsilon (n+2)/(4(n+8))$ is the correlation length exponent, and
there is no correction to the prefactor at order $\epsilon$. 

To obtain the leading
contribution to ${\cal A}$ we have to go to order $g^2$, where we obtain
\begin{eqnarray}
{\cal A} = 1 + \frac{n+2}{144 \epsilon} g^2 + &&\frac{n+2}{18} \left(
\frac{\mu^{\epsilon} g}{S_{d+1}} \right)^2 \frac{\partial}{\partial p^2}
\int \frac{ d^{d+1} p_1}{ (2 \pi )^{d+1}} \frac{ d^{d+1} p_2}{ (2 \pi)^{d+1}}
\Biggl( \nonumber \\
&&~~~~~~~~\left. \frac{1}{(p_1^2 + \Delta_+^2) ( p_2^2 + \Delta_+^2) ( (p + p_1 +p_2)^2
+ \Delta_+^2)} \Biggr) \right|_{p^2 = - \Delta_+^2}
\end{eqnarray}
The integral can be performed by transforming to the usual parametric representation,
which yields
\begin{equation}
{\cal A} = 1 + \frac{n+2}{144 \epsilon} g^2 - \frac{n+2}{18} g^2 \left(
\frac{\mu}{\Delta_+} \right)^{2 \epsilon} \frac{\Gamma^2 (2-\epsilon/2)
\Gamma(\epsilon)}{4} \int_0^1 dx \int_0^1 dy 
\frac{(1-y) y^{-\epsilon/2} x^{\epsilon/2} (1-x)^{\epsilon/2}}{
(1 - x (1-x) (1-y))^{\epsilon}}
\end{equation}
Evaluating the integral as a power series in $\epsilon$, we find that the poles in
$\epsilon$ cancel. Finally, replacing $g \rightarrow g^{\ast}$, we find
\begin{equation}
{\cal A} = \left(\frac{\Delta_+}{\mu}\right)^{\eta} \left( 1 + 0.2823615146
\frac{n+2}{(n+8)^2}
\epsilon^2 \right)
\end{equation}
where the exponent $\eta = (n+2) \epsilon^2 / (2 (n+8)^2 )$.
 
\subsection{$t<0$}
First we determine the value of $N_0 = \langle \phi \rangle$. Ordinary
bare perturbation theory gives
\begin{equation}
N_0 = \sqrt{ \frac{6 |t_0 |}{u_0} } \left[ 
1 - \frac{u_0}{4 |t_0|} \int \frac{d^{d+1} p}{(2 \pi)^{d+1}}
\left( \frac{1}{p^2 + 2 |t_0|} - \frac{1}{p^2} \right) \right]
\label{baren0}
\end{equation}
Re-expressing in terms of the renormalized $t$ and $g$, and evaluating at
$g=g^{\ast}$, we find (we can ignore the wavefunction renormalization 
$Z$ at this order)
\begin{equation}
N_0 = \mu^{1-\epsilon/2} \sqrt{ \frac{n+8}{2\epsilon}} \left( \frac{2|t|}{\mu^2}
\right)^{\beta}
\label{valn0}
\end{equation}
where the exponent $\beta = 1/2 - 3 \epsilon/(2 (n+8))$.

To get the energy scale measuring deviation from criticality, we consider the cases
$n=1$ and $n\geq 2$ separately:
\subsubsection{$n=1$}
Bare perturbation theory tells us that $\chi (p)$ is given by is
\begin{eqnarray}
\chi^{-1} (p) = p^2 + t_0 + \frac{u_0 N_0^2}{2} - && \frac{u_0^2  N_0^2}{2} \int
\frac{d^{d+1} p_1}{(2
\pi)^{d+1}} \frac{1}{((p+p_1)^2 + 2 |t_0| )(p_1^2 + 2 |t_0 |)}
\nonumber
\\ &&~~~~~~~~~~~~~~+ \frac{u_0}{2}
\int \frac{d^{d+1} p_1}{(2
\pi)^{d+1}} \left(\frac{1}{p_1^2 + 2 |t_0|} - \frac{1}{p_1^2}\right)
\end{eqnarray}
Using (\ref{baren0}), 
re-expressing in terms of the renormalized $t$ and $g$, and
evaluating at
$g=g^{\ast}$, we find (again ignoring  
$Z$ at this order) the energy gap $\Delta_-$ by solving $\chi^{-1} ( p^2 = -\Delta_-^2)
= 0$:
\begin{equation}
\Delta_-^2 = \mu^2 \left( 1 + \frac{\pi\sqrt{3}-3}{6} \epsilon \right)
\left( \frac{2 |t|}{\mu^2} \right)^{2 \nu}
\label{valdelm}
\end{equation}

\subsubsection{$n\geq 2$}
In this case we will use the stiffness, $\rho_s$ as a measure of deviation from
criticality. We compute the transverse susceptibility (measured in a direction
orthogonal to the condensate) in bare perturbation theory:
\begin{eqnarray}
\chi^{-1}_{\perp} (p) = p^2 + t_0 + \frac{u_0 N_0^2}{6} - && \frac{u_0^2  N_0^2}{9}
\int
\frac{d^{d+1} p_1}{(2
\pi)^{d+1}} \frac{1}{(p+p_1)^2 (p_1^2 + 2|t_0|)}
\nonumber
\\ &&~~~~~~~~~~~~~~+ \frac{u_0}{6}
\int \frac{d^{d+1} p_1}{(2
\pi)^{d+1}} \left(\frac{1}{p_1^2 + 2|t_0|} - \frac{1}{p_1^2}\right)
\end{eqnarray}
Again, we use (\ref{baren0}), 
re-express in terms of the renormalized $t$ and $g$, and
evaluate at
$g=g^{\ast}$, to obtain:
\begin{equation}
\chi^{-1}_{\perp} (p) = p^2 + \frac{2 \epsilon |t|}{n+8} \left[
\left(1 + \frac{2|t|}{p^2} \right) \log \left( 1 + \frac{p^2}{2|t|} \right) - 1
\right]
\end{equation}
In the region $p^2 \ll |t|$, the above result takes the simple form
\begin{equation}
\chi^{-1}_{\perp} (p) = p^2 \left( 1 + \frac{\epsilon}{2(n+8)} \right).
\end{equation}
We can therefore identify
\begin{equation}
\frac{N_0^2}{\rho_s} = \left( 1 - \frac{\epsilon}{2(n+8)} \right)
\left( \frac{2|t|}{\mu^2} \right)^{\eta\nu}
\label{valrhos}
\end{equation}

\subsection{Universal ratios}
For completeness, we list here the universal ratios that can be constructed out of
the $t>0$ and $t<0$ results. For $n=1$, we can take the ratios of the gaps
$\Delta_-$ and $\Delta_+$
\begin{equation}
\frac{\Delta_-}{\Delta_+} = 2^{\nu} \left( 1 + \frac{\pi\sqrt{3}-3}{12} \epsilon
\right) 
\end{equation}
The analog of this ratio for $n\geq 2 $ is 
\begin{equation}
\frac{\rho_s}{\Delta_+^{d-1}} = 2^{(d-1)\nu} \left( \frac{n+8}{2 \epsilon} \right)
\left( 1 + \frac{\epsilon}{2(n+8)} \right)
\end{equation}
A second set of ratios emerges from the ratios of the field scale. Now we
have for $n=1$
\begin{equation}
\frac{N_0^2}{\Delta_-^{d-1} {\cal A}} = \frac{n+8}{2 \epsilon} 
\left( 1 - \frac{\pi \sqrt{3} - 3}{6} \epsilon \right),
\end{equation}
and for $n\geq 2$
\begin{equation}
\frac{N_0^2}{\rho_s {\cal A}} = 1 - \frac{\epsilon}{2(n+8)}.
\end{equation}

\section{Computations for ${\cal S}$ for $T>0$}
\label{Tgt0}
This appendix will present formal results for the system ${\cal S}$ (Eqn (\ref{cals})
to order $u_0^2$ at nonzero $T$. In Section~\ref{sec:method}, we outlined how to compute
these using a two-step process: ({\em i\/}) obtain an effective action
${\cal S}_{eff}$ (Eqn (\ref{seff})) for the $\omega_n=0$ mode; ({\em ii\/}) compute of
correlations of observables under
${\cal S}_{eff}$. 

First, for future use, let us obtain the value of the mass subtraction $m_{0c}^2$.
Consider the susceptibility of the theory in ${\cal S}_{eff}$,
but with bare mass $m_0^2$; to order $u_0^2$, this is given in bare perturbation theory 
at $T=0$ by
\begin{eqnarray}
&&\frac{1}{Z} \chi^{-1} (p) =  p^2 + m_0^2 + u_0 \left( 
\frac{n+2}{6} \right)  \int \frac{d^{d+1} p_1}{(2\pi)^{d+1}}
\frac{1}{p_1^2 + m_0^2} \nonumber \\
&&~~~~~- u_0^2 \left(\frac{n+2}{6}\right)^2 
\left(\int \frac{d^{d+1} p_1}{(2\pi)^{d+1}}
\frac{1}{ p_1^2 + m_0^2}\right)
\left( \int \frac{d^{d+1} p_2}{(2\pi)^{d+1}}
\frac{1}{(p_2^2 + m_0^2)^2}\right) \nonumber \\
&&~~~~~- u_0^2 \left( \frac{n+2}{18} \right)  
\int
\frac{d^{d+1} p_1}{(2\pi)^{d+1}} \frac{d^{d+1} p_2}{(2\pi)^{d+1}}
\frac{1}{(p_1^2 + m_0^2)(p_2^2 + m_0^2)((p-p_1 -p_2)^2 + m_0^2)}
\label{chim0}
\end{eqnarray}
The critical point is determined by the value $m_0 = m_{0c}$ at which
$\chi^{-1}(p=0) = 0$ at $T=0$. Solving (\ref{chim0}) for this condition order by order in
$u_0$ we obtain
\begin{equation}
m_{0c}^2 = - u_0 \left( \frac{n+2}{6}\right) \int \frac{d^{d+1}p}{(2 \pi)^{d+1}}
\frac{1}{p^2} + u_0^2 \left( \frac{n+2}{18} \right)
\int \frac{d^{d+1}p_1}{(2 \pi)^{d+1}} \frac{d^{d+1}p_2}{(2 \pi)^{d+1}}
\frac{1}{p_1^2 p_2^2 (p_1 + p_2)^2}
\label{valm0c}
\end{equation}
In subsequent computations,  in all propagators carrying a non-zero
frequency, we will insert the mass
\begin{equation}
m_0^2 = m_{0c}^2 + t_0,
\label{massnz}
\end{equation}
with $m_{0c}^2$ given by 
(\ref{valm0c}), and expand in powers of $u_0$: all non-zero frequency propagators in
the resulting expression will therefore have mass $t_0$. 

Let us now obtain the couplings in the effective action ${\cal S}_{eff}$ for the
$\omega_n=0$ mode to order $u_0^2$. 
By ordinary perturbation theory in the finite frequency modes we obtain
\begin{eqnarray}
&&\widetilde{C}_2 (k) = Z k^2 + Z t_0 + u_0 \left( 
\frac{n+2}{6} \right) \left[ T\sum_{\epsilon_n \neq 0} \int \frac{d^d q}{(2\pi)^d}
\frac{1}{\epsilon_n^2 + q^2 + t_0} - \int
\frac{d^{d+1}p}{(2\pi)^{d+1}} \frac{1}{p^2}\right] \nonumber \\
&&~~~~~- u_0^2 \left(\frac{n+2}{6}\right)^2 
\left[T\sum_{\epsilon_n \neq 0} \int \frac{d^d q_1}{(2\pi)^d}
\frac{1}{\epsilon_n^2 + q_1^2 + t_0} - \int
\frac{d^{d+1}p}{(2\pi)^{d+1}} \frac{1}{p^2} \right]
\left(T\sum_{\Omega_n \neq 0} \int \frac{d^d q_2}{(2\pi)^d}
\frac{1}{(\Omega_n^2 + q_2^2 + t_0)^2}\right) \nonumber \\
&&~~~~~- u_0^2 \left( \frac{n+2}{18} \right) \left[ T^2\sum_{\epsilon_n\neq 0,
\Omega_n \neq 0, \epsilon_n + \Omega_n \neq 0} 
\int
\frac{d^d q_1}{(2\pi)^d} \frac{d^d q_2}{(2\pi)^d}
\frac{1}{(\epsilon_n^2 + q_1^2 + t_0)(\Omega_n^2 + q_2^2 +
t_0)} \right.\nonumber \\
&& \left.~~~~~~~~~~~~~~~~\times \frac{1}{((\epsilon_n +
\Omega_n)^2 + (k-q_1 -q_2)^2 +t_0)} -  
\int \frac{d^{d+1}p_1}{(2 \pi)^{d+1}} \frac{d^{d+1}p_2}{(2 \pi)^{d+1}}
\frac{1}{p_1^2 p_2^2 (p_1 + p_2)^2}
\right].
\label{resC2t}
\end{eqnarray}
We have implicitly assumed above that $Z = 1 + {\cal O}(g^2)$ and only written $Z$
where it is need for the order $g^2$ result; we will continue to do this in the
remainder of the appendix. In a similar manner, we can obtain the value of
$\widetilde{C}_4$:
\begin{eqnarray}
&&\widetilde{C}_4 (k_1 , k_2 , k_3, -k_1-k_2-k_3 ) = 
u_0 \nonumber \\
&&~~~~~~~~~~~~~~- u_0^2 \left( \frac{n+8}{6} \right) T \sum_{\epsilon_n \neq 0}
\int \frac{d^d q}{(2 \pi)^d} \mbox{Sym}_k \frac{1}{(\epsilon_n^2 + q^2 + t_0)(\epsilon_n^2
+ (k_1 +k_2 - q)^2 + t_0)},
\label{resC4}
\end{eqnarray}
where the symbol $\mbox{Sym}_k$ denotes that the expression following it has to be
symmetrized among the momenta $k_1$, $k_2$, $k_3$, $k_4 = -k_1 -k_2-k_3$.
All other couplings in ${\cal S}_{eff}$ are zero at order $u_0^2$.

We now perform the renormalizations of the super-renormalizable classical theory to obtain
$C_2$ and $C_4$. First, to order $u_0^2$, it is easy to see that $C_4 = \widetilde{C}_4$.
For $\widetilde{C}_2$, in addition to the tadpole renormalization in (\ref{tadpole}) there is a
two-loop renormalization that has to be included for $d$ close to 3
\begin{eqnarray}
C_2 (k) = && \widetilde{C}_2 (k) + \left( \frac{n+2}{6} \right) \int \frac{d^d k_1 }{(2
\pi)^d}
\frac{\widetilde{C}_4 (k, -k, k_1, -k_1 )}{k_1^2} \nonumber \\
&&~~~~-\left( \frac{n+2}{18} \right) \int \frac{d^d k_1 }{(2 \pi)^d} \frac{d^d k_2 }{(2
\pi)^d} \frac{\widetilde{C}_4 (k, k_1, k_2, -k-k_1-k_2) \widetilde{C}_4 (-k, -k_1,-k_2,k+k_1+k+2)}{
(k_1^2 + T^2 ) (k_2^2 + T^2) ( (k+k_1+k_2)^2 + T^2)}
\label{pauliv}
\end{eqnarray}
This, in fact, completes the set of renormalizations, and there are no new terms that have
to be accounted for at higher orders in $u_0$ in the super-renormalizable classical theory.
To avoid an infrared divergence in $d=3$, we have performed the two-loop renormalization
above at an arbitrarily chosen Pauli-Villars mass equal to $T$. Our results for the coupling
constants will therefore depend upon this choice of renormalization scheme, but all physical
observables will be independent of it. Combining (\ref{resC2t}) and (\ref{pauliv}) we get
\begin{eqnarray}
&&{C}_2 (k) = Z k^2 + Z t_0 + u_0 \left( 
\frac{n+2}{6} \right) \left[ T\sum_{\epsilon_n} \int \frac{d^d q}{(2\pi)^d}
\frac{1}{ q^2 + \sigma(\epsilon_n)} - \int
\frac{d^{d+1}p}{(2\pi)^{d+1}} \frac{1}{p^2}\right] \nonumber \\
&&~~~~~- u_0^2 \left(\frac{n+2}{6}\right)^2 
\left[T\sum_{\epsilon_n} \int \frac{d^d q_1}{(2\pi)^d}
\frac{1}{ q_1^2 + \sigma (\epsilon_n) } - \int
\frac{d^{d+1}p}{(2\pi)^{d+1}} \frac{1}{p^2} \right]
\left(T\sum_{\Omega_n \neq 0} \int \frac{d^d q_2}{(2\pi)^d}
\frac{1}{(\Omega_n^2 + q_2^2 + t_0)^2}\right) \nonumber \\
&&~~~~~- u_0^2 \left( \frac{n+2}{18} \right) \left[ T^2\sum_{\epsilon_n,
\Omega_n } 
\int
\frac{d^d q_1}{(2\pi)^d} \frac{d^d q_2}{(2\pi)^d}
\frac{1}{( q_1^2 + \sigma(\epsilon_n))( q_2^2 +
\sigma(\Omega_n))((k-q_1 -q_2)^2 +\sigma(\epsilon_n + \Omega_n))} \right. \nonumber \\
&&~~~~~~~~~~~~~~~~~~~~~~~~~~~~~~~~~~~~~~~~~~~~~~~~~~~~~~~~~~~~~~-  
\int \frac{d^{d+1}p_1}{(2 \pi)^{d+1}} \frac{d^{d+1}p_2}{(2 \pi)^{d+1}}
\frac{1}{p_1^2 p_2^2 (p_1 + p_2)^2}
\Biggr] \nonumber \\
&&~- u_0^2 \left( \frac{n+2}{18} \right)  T^2
\int \frac{d^d q_1}{(2\pi)^d} \frac{d^d q_2}{(2\pi)^d} \left[
\frac{1}{( q_1^2 + T^2)(q_2^2 +
T^2)((k-q_1 -q_2)^2 +T^2 )} - \frac{1}{ q_1^2 q_2^2 (k-q_1 -q_2)^2 }\right]
\label{resC2}
\end{eqnarray}
where we have defined 
\begin{equation}
\sigma (\epsilon_n) \equiv \epsilon_n^2 + t_0 ( 1 - \delta_{\epsilon_n , 0} ).
\end{equation}
The values of the couplings $R$, $K$, and $U$ now follow directly from (\ref{defu}) and
the results (\ref{resC4}), (\ref{resC2}) above.

Let us now apply the above procedure to obtain the perturbative result for the
dynamic susceptibility $\chi (k , i\omega_n )$ at finite $T$. As discussed in
Section~\ref{sec:method}, this result will be valid everywhere in the phase diagram
of Fig~\ref{f1}, except in the shaded region $|T - T_c (t) | \ll T_c (t)$.
The simplest way to proceed is to introduce an external source term coupling
to the field $\phi$ and then to proceed in the two-step procedure noted at
the beginning of this Appendix. We omit the details and state the final result
\begin{eqnarray}
&&  \chi^{-1} (k, i\omega_n ) = 
Z k^2 + Z \omega_n^2 + R 
 - u_0 \left( 
\frac{n+2}{6} \right) T \int
\frac{d^{d}q}{(2\pi)^{d}} \frac{R}{q^2 (q^2 + R)} \nonumber \\
&&~~~~~+ u_0^2 \left(\frac{n+2}{6}\right)^2 
\left[T \int \frac{d^d q_1}{(2\pi)^d}
\frac{R}{ q_1^2 (q_1^2 + R)}\right]
\left(T\sum_{\Omega_n} \int \frac{d^d q_2}{(2\pi)^d}
\frac{1}{(q_2^2 + \widetilde{\sigma} (\Omega_n))^2}\right) \nonumber \\
&&~~~~~- u_0^2 \left( \frac{n+2}{18} \right)  T^2\sum_{\epsilon_n,\Omega_n} 
\int
\frac{d^d q_1}{(2\pi)^d} \frac{d^d q_2}{(2\pi)^d}
\left[ \frac{1}{(q_1^2 + \widetilde{\sigma}(\epsilon_n))(q_2^2 +
\widetilde{\sigma}(\Omega_n))((k-q_1 -q_2)^2 +
\widetilde{\sigma}(\omega_n - \epsilon_n - \Omega_n))} \right. \nonumber \\
&&~~~~~~~~~~~~~~~~~~~~~~~~~~~~~~~~~~~~~~~~~~~~~~~~~~~~~~~
 \left. -~~\frac{1}{(q_1^2 + \sigma(\epsilon_n))(q_2^2 +
\sigma(\Omega_n))((q_1 +q_2)^2 +
\sigma(\epsilon_n + \Omega_n))} 
\right] \nonumber \\
&&~~~~~+ u_0^2 \left( \frac{n+2}{18} \right)  T^2
\int \frac{d^d q_1}{(2\pi)^d} \frac{d^d q_2}{(2\pi)^d} \left[
\frac{1}{( q_1^2 + T^2)(q_2^2 +
T^2)((q_1 +q_2)^2 +T^2 )} - \frac{1}{ q_1^2 q_2^2 (q_1 +q_2)^2 }\right],
\label{reschi}
\end{eqnarray}
where 
\begin{equation}
\widetilde{\sigma}(\epsilon_n) \equiv \epsilon_n^2 + R \delta_{\epsilon_n, 0} + 
t_0 ( 1 - \delta_{\epsilon_n ,
0} ).
\label{restsigma}
\end{equation}
Notice that in (\ref{reschi}) 
we do not expand out the $u_0$ dependent expression for $R$
given in (\ref{resR}), but instead treat
$R$ as variable formally independent of $u_0$; this is required by the method of
Section~\ref{sec:method}.

The equations (\ref{resC2}) and (\ref{reschi}), are the main
results of this appendix, and will be used in the body of the paper.

In the following subsections of this appendix, we will evaluate the formal results above to obtain
explicit two-loop expressions for some quantities at $t=0$. Our main purpose in doing this is to
demonstrate the consistency of our approach, by explicitly displaying the cancellation of all
ultraviolet and infrared divergences and the collapse of the results into the scaling forms of
Sec.~\ref{intro}.

\subsection{Evaluation of $R$}
\label{evalR}
As noted above, we will restrict our results to the critical coupling $t=0$.
We begin with (\ref{resC2}) and the definition (\ref{defu}). Explicitly evaluating out the
one-loop contributions in terms of the functions introduced in Section~\ref{sec:cc} (and identity
(\ref{id1})),
and expressing in terms of the coupling $g$ using (\ref{defg}), we find to order $g^2$
\begin{eqnarray}
&& R = g T^2 \left( \frac{\mu}{T} \right)^{\epsilon} 
\left( \frac{n+2}{6} \right) \left( 1+
\frac{n+8}{6
\epsilon} g
\right) F_d (0) - g^2 T^2 \left( \frac{\mu}{T} \right)^{2\epsilon} 
\left( \frac{n+2}{6} \right)^2 \left( \frac{1}{\epsilon} - G^{\prime} (0) \right) F_d (0)
\nonumber \\
&&~~~~~- u_0^2 \left( \frac{n+2}{18} \right) \left[ T^2\sum_{\epsilon_n,
\Omega_n } 
\int
\frac{d^d q_1}{(2\pi)^d} \frac{d^d q_2}{(2\pi)^d}
\frac{1}{( \epsilon_n^2 + q_1^2 )(\Omega_n^2 +  q_2^2 )((\epsilon_n + \Omega_n)^2 +
(q_1 + q_2)^2)} \right. \nonumber \\
&&~~~~~~~~~~~~~~~~~~~~~~~~~~~~~~~~~~~~~~~~~~~~~~~~~~~~~~~~~~~~~~-  
\int \frac{d^{d+1}p_1}{(2 \pi)^{d+1}} \frac{d^{d+1}p_2}{(2 \pi)^{d+1}}
\frac{1}{p_1^2 p_2^2 (p_1 + p_2)^2}
\Biggr] \nonumber \\
&&~~~~~- u_0^2 \left( \frac{n+2}{18} \right)  T^2
\int \frac{d^d q_1}{(2\pi)^d} \frac{d^d q_2}{(2\pi)^d} \left[
\frac{1}{( q_1^2 + T^2)(q_2^2 +
T^2)((q_1 +q_2)^2 +T^2 )} - \frac{1}{ q_1^2 q_2^2 (q_1 + q_2)^2 }\right]
\label{aR1}
\end{eqnarray}
Notice that the above expression has poles in $\epsilon$ multiplying the thermal function $F_d
(0)$. Consistency requires that these poles must cancel divergences coming out of the two-loop
frequency summation left unevaluated in (\ref{aR1}). This is indeed what happens. We can see this
by adding and subtracting the following expression to (\ref{aR1}):
\begin{eqnarray}
&& g^2 \mu^{2\epsilon} 
\left( \frac{n+2}{24} \right) \left( \frac{1}{S_{d+1}} \int \frac{d^d q_1}{(2 \pi)^d}
\frac{1}{q_1} \frac{1}{e^{q_1 /T} - 1} \right) \left( \frac{1}{S_{d+1}} \int \frac{d^d q_2}{(2
\pi)^d}
\frac{1}{(q_2^2 + T^2)^{3/2}} \right)  \nonumber \\
&&~~~~~~~~~~= g^2 T^2 \left( \frac{\mu}{T} \right)^{2\epsilon} 
\left( \frac{n+2}{24} \right)  F_d (0)  2 \Gamma(2-\epsilon/2) \Gamma(\epsilon/2)
\label{aR2}
\end{eqnarray}
We absorb the left-hand-side of (\ref{aR2}) into the unevaluated integrals in (\ref{aR1}).
The right-hand-side of (\ref{aR2}) has poles in $\epsilon$ which precisely cancel the poles in
(\ref{aR1}). Setting $g=g^{\ast}$ (Eqn (\ref{valgast})) and expanding in powers of
$\epsilon$, one finds that the
$\mu$ dependence of (\ref{aR1}) also disappears; in this manner we obtain
\begin{equation}
R = \frac{(n+2) \epsilon T^2}{n+8} \left[ 1 + \left( \frac{n+2}{n+8} G^{\prime} (0)
- \frac{n^2 - 8n - 68}{2 (n+8)^2} \right) \epsilon \right] F_d (0) 
- \frac{2 (n+2)\epsilon^2 T^2}{(n+8)^2} I_1
\label{aR3}
\end{equation}
where the number $I$ arises from the frequency summations in (\ref{aR1}) combined with
(\ref{aR2}); evaluating these summations we find
\begin{eqnarray}
&& I_1 = \frac{3}{8 \pi^2} \int d^3 q_1 d^3 q_2 \left[
\frac{n(q_1) ( 1 + n(q_2) + n(|q_1 + q_2|)) - n(q_2 ) n(|q_1 + q_2|)}{q_1 q_2 |q_1 + q_2|
(q_2 + |q_1 + q_2| - q_1 )} \right. \nonumber \\
&&~~~~~~~~~~~~~~~~~~~~~~~~~~~~~~~~~
+ \frac{n(q_2) ( 1 + n(q_1) + n(|q_1 + q_2|)) - n(q_1 ) n(|q_1 +
q_2|)}{q_1 q_2 |q_1 + q_2| (q_1 + |q_1 + q_2| - q_2 )}  \nonumber \\
&& ~~~~~~~~~~~~~~~~~~~~~~~~~~~~~~~~+ \frac{n(q_1)  + n(q_2)  n(|q_1 + q_2|)}{q_1 q_2 |q_1 + q_2|
(q_1 + q_2 + |q_1 + q_2|)}
+ \frac{n(q_2)  + n(q_1)  n(|q_1 + q_2|)}{q_1 q_2 |q_1 + q_2|
(q_1 + q_2 + |q_1 + q_2|)} \nonumber \\
&&~~\left.- \frac{n(q_1)}{q_1} \frac{1}{(q_2^2 + 1)^{3/2}} 
- \frac{n(q_2)}{q_2} \frac{1}{(q_1^2 + 1)^{3/2}} 
+ \frac{8}{3 (q_1^2 + 1)(q_2^2 + 1) ((q_1 + q_2)^2 + 1)}
- \frac{8}{3 q_1^2 q_2^2 (q_1 + q_2)^2 }\right]
\label{aR4}
\end{eqnarray}
where $n(q) \equiv 1/(e^q - 1)$ is the Bose function at unit temperature.
It can be checked by a straightforward asymptotic analysis that the combined integrals in 
(\ref{aR4}) are free of both ultraviolet and infrared divergences; we evaluated the integrals
numerically and found
\begin{equation}
I_1 \approx -15.2
\label{aR5}
\end{equation} 
We also quote the values of the other constants in (\ref{aR3}):
\begin{eqnarray}
G^{\prime} (0) &=& 2.45380858207\ldots\nonumber \\
F_d (0) &=& \frac{2 \pi^2}{3} - \frac{\pi^2 (1 + 2  \ln 2 - 2 \gamma) + 12 \zeta^{\prime} (2)}{3}
\epsilon + {\cal O}(\epsilon^2),
\end{eqnarray}
where $\gamma = 0.577216\ldots$ is Euler's constant, and $\zeta(s)$ is the Reimann zeta function.

\subsection{Evaluation of $\chi (0,0)$}
We can obtain an expression for the static susceptibility, $\chi(0,0)$, at $t=0$
directly from (\ref{reschi})
\begin{eqnarray}
&&  \chi^{-1} (0,0 ) = 
 R 
 - u_0 \left( 
\frac{n+2}{6} \right) T \int
\frac{d^{d}q}{(2\pi)^{d}} \frac{R}{q^2 (q^2 + R)} \nonumber \\
&&~~~~~+ u_0^2 \left(\frac{n+2}{6}\right)^2 
\left[T \int \frac{d^d q_1}{(2\pi)^d}
\frac{R}{ q_1^2 (q_1^2 + R)}\right]
\left[T \int \frac{d^d q_2}{(2\pi)^d}\left( \sum_{\Omega_n \neq 0}
\frac{1}{(\Omega_n^2 + q_2^2)^2} + \frac{1}{(q_2^2 + R)^2}\right) \right]\nonumber \\
&&~~~~~+ 3 u_0^2 \left( \frac{n+2}{18} \right)  T^2\sum_{\Omega_n \neq 0} 
\int
\frac{d^d q_1}{(2\pi)^d} \frac{d^d q_2}{(2\pi)^d}
 \frac{R}{q_1^2 (q_1^2 + R)(\Omega_n^2 + q_2^2 )(\Omega_n^2 + (q_1 +q_2)^2)} 
\nonumber \\ 
&&~~~~~- u_0^2 \left( \frac{n+2}{18} \right)  T^2
\int \frac{d^d q_1}{(2\pi)^d} \frac{d^d q_2}{(2\pi)^d} \left[
\frac{1}{( q_1^2 + R)(q_2^2 + R)((q_1 +q_2)^2 + R )} \right. \nonumber \\
&&\left. ~~~~~~~~~~~~~~~~~~~~~~~~~~~~~~~~~~~~~~~~~~~~~~~- \frac{1}{ (q_1^2 +
T^2)( q_2^2 + T^2) ((q_1 + q_2)^2 + T^2) }\right],
\label{chi1}
\end{eqnarray}
Expressing in terms of $g$ using (\ref{defg}), and rearranging terms a bit, we obtain
\begin{eqnarray}
\chi^{-1}(0,0) = && R - g \mu^{\epsilon} T  \left( \frac{n+2}{6} \right) \left( 1 +
\frac{n+8}{6 \epsilon}g \right)
\left[ \frac{1}{S_{d+1}} \int \frac{d^{d}q_1 }{(2\pi)^{d}} \frac{R}{q_1^2 (q_1^2 + R)}\right]
\nonumber \\
&&~~~~~~~~~~~~~~~\times
\left[1 - g \mu^{\epsilon} T \left( \frac{n+8}{6} \right) \frac{1}{S_{d+1}}
\int
\frac{d^d q_2}{(2\pi)^d} \left( \sum_{\Omega_n \neq 0}
\frac{1}{(\Omega_n^2 + q_2^2)^2} + \frac{1}{(q_2^2 + R)^2} \right) \right] \nonumber \\
&&~~~~~~~~~~~~~~~~~~~~~~~~~~~~~~~~~ + g^2
\mu^{2
\epsilon} T^2 
\left( \frac{n+2}{18} \right) \left(I_2 + I_3 \right)
\label{chi2}
\end{eqnarray}
where
\begin{equation}
I_2 = \frac{3}{S_{d+1}^2} \sum_{\Omega_n \neq 0} 
\int
\frac{d^d q_1}{(2\pi)^d} \frac{d^d q_2}{(2\pi)^d}
 \frac{R}{q_1^2 (q_1^2 + R)(\Omega_n^2 + q_2^2 )} \left( \frac{1}{\Omega_n^2 + (q_1 +q_2)^2}
- \frac{1}{\Omega_n^2 + q_2^2} \right)
\label{vali2} 
\end{equation}
and
\begin{equation}
I_3 = - \frac{1}{S_{d+1}^2} \int \frac{d^d q_1}{(2\pi)^d} \frac{d^d q_2}{(2\pi)^d} \left[
\frac{1}{( q_1^2 + R)(q_2^2 + R)((q_1 +q_2)^2 + R )} - \frac{1}{ (q_1^2 + T^2)( q_2^2 + T^2)
((q_1 + q_2)^2 + T^2) }\right]
\label{vali3}
\end{equation}

Let us now evaluate some of the integrals in (\ref{chi2}) to the needed
accuracy in $\epsilon$. First, we have
\begin{eqnarray}
\frac{1}{S_{d+1}}\int \frac{d^{d}q_1 }{(2\pi)^{d}} \frac{R}{q_1^2 (q_1^2 + R)} &=& 
2 \pi R^{(1-\epsilon)/2} \left( 1 + \frac{1 - 2 \ln 2}{2} \epsilon + \ldots \right)
\nonumber \\
\frac{1}{S_{d+1}}\int \frac{d^{d}q_1 }{(2\pi)^{d}} \frac{1}{(q_1^2 + R)^2} &=&
\pi R^{-(1+\epsilon)/2} \nonumber \\
\frac{1}{S_{d+1}}\int \frac{d^{d}q_1 }{(2\pi)^{d}} \sum_{\Omega_n \neq
0} \frac{1}{(\Omega_n^2 + q_1^2)^2} &=&
T^{-1-\epsilon} \left( \frac{1}{\epsilon} - G^{\prime} (0) \right). 
\label{chi3}
\end{eqnarray}
where the last equation is related to (\ref{id1}).
The integrals over momenta in (\ref{vali2}) can be performed exactly in $d=3$ (which is all we
need) and give
\begin{equation}
I_2 = 12 \pi^2 \sum_{\Omega_n \neq 0} \left[ \ln \left(1 + \frac{\sqrt{R}}{2 |\Omega_n|} \right)
- \frac{\sqrt{R}}{2 |\Omega_n|} \right]
\label{chi4}
\end{equation}
Now notice that $ R \sim \epsilon T^2 \ll T^2$. In this limit we can get the leading result for
$I_2$ simply by expanding (\ref{chi4}) to leading order in $R$; this leads to
\begin{eqnarray}
I_2 &=& - \frac{3R}{4 T^2} \sum_{n=1}^{\infty} \frac{1}{n^2} \nonumber \\
&=& - \frac{\pi^2 R}{8 T^2} .
\label{chi5}
\end{eqnarray}
 The integral in (\ref{vali3}) can also be evaluated and we find
\begin{equation}
I_3 = 2 \pi^2 \ln (R /T^2). 
\label{chi6}
\end{equation}

We are now ready to assemble all these results into (\ref{chi2}). Expanding (\ref{chi2}) in
powers of $g$ to order $g^2$ one finds, as expected, that all the poles in $\epsilon$
cancel. Setting $g = g^{\ast}$ (Eqn (\ref{valgast})) and expanding in powers of $\epsilon$
(while keeping $R$ fixed),
one finds that all the $\mu$ dependence disappears and the resulting expression takes the
form
\begin{eqnarray}
\chi^{-1} (0,0) = R - \epsilon \left( \frac{n+2}{n+8} \right) && 2 \pi T \sqrt{R} 
\left[ 1 + \epsilon \left( \frac{20 +2n - n^2}{2 (n+8)^2} + \frac{1 - 2 \ln 2}{2}
- \frac{1}{2} \ln \frac{R}{T^2} + G^{\prime} (0) \right) \right] \nonumber \\
&& + \epsilon^2 \left( \frac{n+2}{n+8} \right) 2 \pi^2 T^2 \left( 1 + \frac{2}{n+8} \ln
\frac{R}{T^2} \right).
\label{chi7}
\end{eqnarray} 
We have retained all terms, which after inserting (\ref{aR3}), are of order $\epsilon^{5/2}$
or larger. The three loop corrections, of order $u_0^3$, contain a contribution like
$\epsilon^3 T /\sqrt{R} \sim \epsilon^{5/2}$, so (\ref{chi7}) does not contain all 
terms of order $\epsilon^{5/2}$; it does, however, include all terms of 
order $\epsilon^2 \ln \epsilon$ or smaller.

\subsection{Evaluation of $\partial \chi^{-1} (k,0)/\partial k^2 |_{k=0}$}

From (\ref{reschi}) we have at $t=0$,
\begin{eqnarray}
\left. \frac{\partial \chi^{-1} (k,0)}{\partial k^2} \right|_{k=0} = &&
1 - \frac{n+2}{144 \epsilon} g^2 - \frac{n+2}{18} \left(
\frac{\mu^{\epsilon} g T}{S_{d+1}} \right)^2 \frac{\partial}{\partial k^2}
\int \frac{d^d q_1}{(2 \pi)^d} \frac{d^d q_2}{(2 \pi)^d}  \left(
\sum_{\epsilon_n, \Omega_n} \right.
\nonumber \\
&&~~~~~~~~~~~~~~~~~~~~~~
\left. \left.
\frac{1}{(q_1^2 + \widetilde{\sigma}(\epsilon_n))(q_2^2 +
\widetilde{\sigma}(\Omega_n))((k-q_1 -q_2)^2 + \widetilde{\sigma}(\omega_n + \Omega_n))}
\right)
\right|_{k=0}
\label{achi1}
\end{eqnarray}
where $\tilde{\sigma} (\epsilon_n)$ is defined in (\ref{restsigma}) with $t_0 = 0$.
Now add and subtract the following integral from the above
\begin{eqnarray}
I_4 &=& \frac{n+2}{18} \left(
\frac{\mu^{\epsilon} g}{S_{d+1}} \right)^2 \frac{\partial}{\partial k^2}
\int \frac{d^{d+1} p_1}{(2 \pi)^{d+1}} \frac{d^{d+1} p_2}{(2 \pi)^{d+1}}
\left. \frac{1}{(p_1^2 + T^2) (p_2^2 + T^2) ((k-p_1 - p_2)^2 + T^2)} \right|_{k=0} \nonumber \\
&=& g^2 \left( \frac{n+2}{18} \right) \left[ - \frac{1}{8 \epsilon} + \frac{1}{4} 
\ln \frac{T}{\mu} + I_5  + {\cal O}(\epsilon) \right],
\label{achi2}
\end{eqnarray}
with the constant $I_5 = 0.22657560322\ldots$.
The poles in $\epsilon$ then cancel, and the remainder of (\ref{achi1}) can be evaluated at
$\epsilon=0$ and $g = g^{\ast}$. 
Let us now define the momentum integral
\begin{eqnarray}
X(a_1, a_2, a_3) &\equiv& \frac{1}{S_4^2} \frac{\partial}{\partial k^2}
\int \frac{d^{3} q_1}{(2 \pi)^{3}} \frac{d^{3} q_2}{(2 \pi)^{3}}
\left. \frac{1}{(q_1^2 + a_1) (q_2^2 + a_2) ((k-q_1 - q_2)^2 + a_3)} \right|_{k=0}
\nonumber \\
&=& - \pi \int_0^1 dx \int_0^1 dy \frac{ (1-y) \sqrt{y x (1-x)}}{
a_1 (1-x) y + a_2 x (1-x) (1-y) + a_3 x y},
\label{achi3}
\end{eqnarray}
where in the second equation we have transformed to the usual parametric representation.
Then we can write (\ref{achi1}) in the form
\begin{eqnarray}
\left. \frac{\partial \chi^{-1} (k,0)}{\partial k^2} \right|_{k=0} = &&
1 - \eta \ln \frac{T}{\mu} - \epsilon^2 \frac{2(n+2)}{(n+8)^2}\left[I_5 +
T^2 \sum_{\epsilon_n, \Omega_n} X(\widetilde{\sigma}(\epsilon_n),
\widetilde{\sigma}(\Omega_n),\widetilde{\sigma}(\epsilon_n+\Omega_n))
\right. \nonumber \\
&&~~~~~~~~~~~~~~~~~~
\left. - \int \frac{d \epsilon}{2 \pi} \frac{d \Omega}{2 \pi}
X(\epsilon^2 + T^2, 
\Omega^2 + T^2, (\epsilon + \Omega)^2 + T^2) 
\right]
\label{achi4}
\end{eqnarray}
It is now not difficult to show that the combination of the summation and integration within
the square brackets in (\ref{achi4}) is free of both ultraviolet and infrared divergences.
In fact, this combination is a dimensionless quantity which is a function only of the
dimensionless ratio $R/T^2$. Now we know from (\ref{aR3}) that $R/T^2 \ll 1$, and in this
limit, the term in the square bracket in (\ref{achi4}) is dominated by the single term
in the summation with $\epsilon_n = \Omega_n = 0$; we have therefore
\begin{eqnarray}
\left. \frac{\partial \chi^{-1} (k,0)}{\partial k^2} \right|_{k=0} &=& 
1 - \eta \ln \frac{T}{\mu} - \epsilon^2 \frac{2(n+2)}{(n+8)^2} \left[
T^2 X(R,R,R) \left(1 + {\cal O}( \sqrt{R/T^2} ) \right) \right] \nonumber \\
&=& \left( \frac{T}{\mu} \right)^{-\eta} \left[
1  + \epsilon^2 \frac{0.5959380965224 (n+2) }{(n+8)^2 } \frac{T^2}{R}
\left(1 + {\cal O}( \sqrt{R/T^2} ) \right)
\right]
\end{eqnarray}

\begin{figure}
\epsfxsize=6in
\centerline{\epsffile{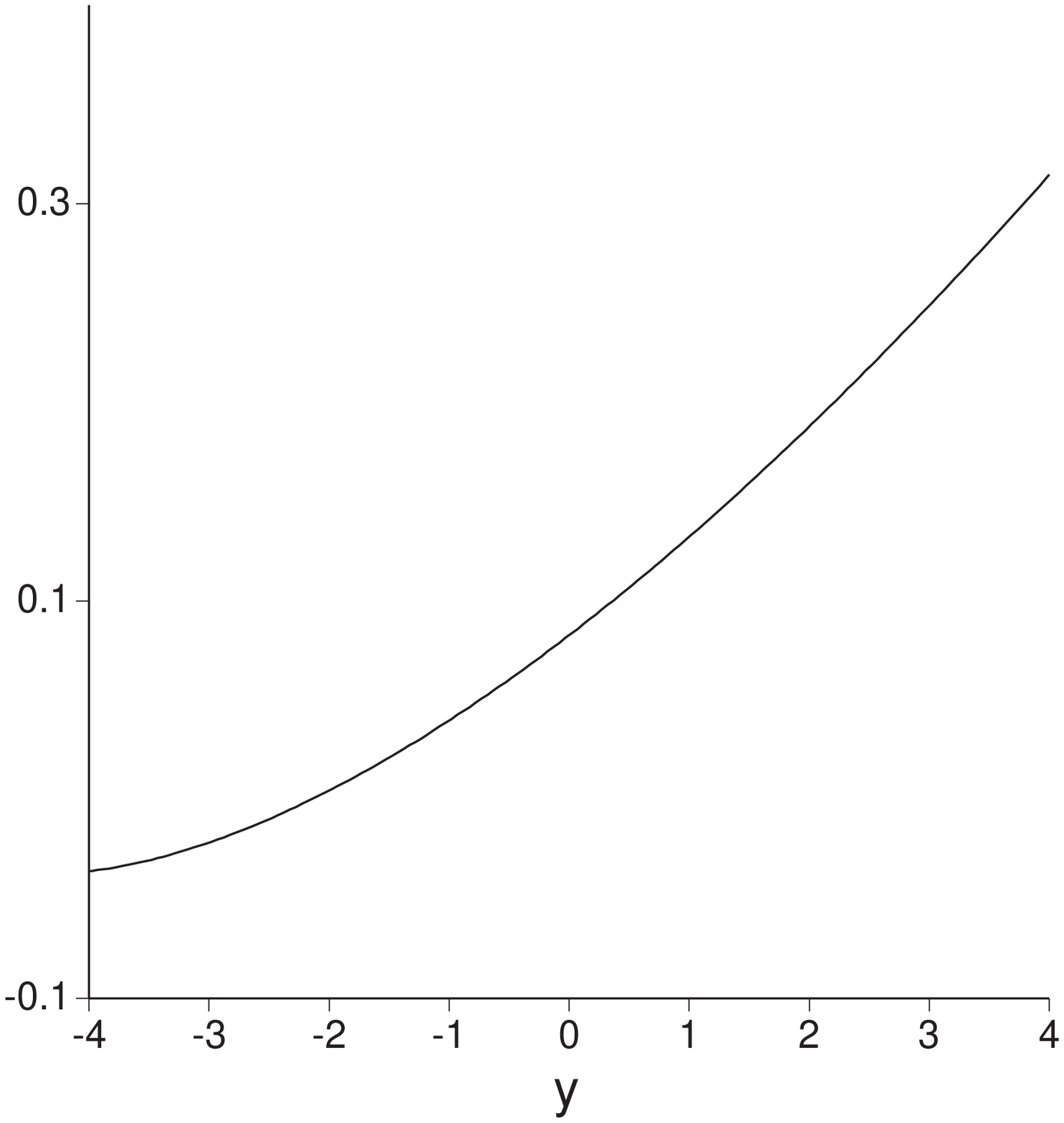}}
\vspace{0.5in}
\caption{A plot of the universal scaling function $\Upsilon (y)$, defined in
(\protect\ref{up10}), for $d=3$. Notice that it is smooth at $y=0$. It is analytic for all real
$y > -2 \pi$.}
\label{f4}
\end{figure}


\begin{references}

\bibitem{hertz} J.A. Hertz, Phys. Rev. B {\bf 14}, 525 (1976).

\bibitem{suzuki} M. Suzuki, Prog. Theor. Phys. {\bf 56}, 1454 (1976).

\bibitem{lawrie} I.D. Lawrie, J. Phys. C {\bf 11}, 1123 (1978); {\em ibid\/} {\bf 11}, 3857
(1978).

\bibitem{otherbose} K.K. Singh, Phys. Rev. B {\bf 12}, 2819 (1975); {\em ibid\/}
{\bf 17}, 324 (1978).

\bibitem{cres} R.J. Creswick and F.W. Wiegel, Phys. Rev. A {\bf 28}, 1579 (1983).

\bibitem{weichmann} M. Rasolt, M.J. Stephen, M.E. Fisher and  P.B. Weichmann,
Phys. Rev. Lett. {\bf 53}, 798 (1984);
P.B. Weichmann, M. Rasolt, M.E. Fisher, and
M.J. Stephen, Phys. Rev. B {\bf 33}, 4632 (1986).

\bibitem{dsf} D.S. Fisher and P.C. Hohenberg, Phys. Rev. B {\bf 37}, 4936
(1988); see also V.N. Popov, {\em Functional Integrals in Quantum Field
Theory and Statistical Physics\/}, D. Reidel (Boston), 1983.

\bibitem{CHN} S. Chakravarty, B.I. Halperin, and D.R. Nelson,
Phys. Rev. B {\bf 39}, 2344 (1989).

\bibitem{CSY} S. Sachdev and J. Ye, Phys. Rev. Lett. {\bf 69}, 2411 (1992);
A.V. Chubukov and S. Sachdev, Phys. Rev. Lett. {\bf 71}, 169  (1993); 
A.V. Chubukov, S. Sachdev and J. Ye, Phys.  Rev. B {\bf 49}, 11919 (1994).

\bibitem{millis} A.J. Millis, Phys. Rev. B {\bf 48}, 7183 (1993).

\bibitem{sokolpines} A. Sokol and D. Pines, Phys. Rev. Lett. {\bf 71}, 2813 (1993).

\bibitem{sss} S. Sachdev, T. Senthil, and R. Shankar, Phys. Rev. B {\bf 50}, 258 (1994).

\bibitem{contin} M.A. Continentino, Phys. Rep. {\bf 239}, 179 (1994).

\bibitem{conserve} S. Sachdev, Z. Phys. B {\bf 94}, 469 (1994).

\bibitem{uz} U. Zulicke and A.J. Millis, Phys. Rev. B {\bf 51}, 8996 (1995).

\bibitem{scs} S. Sachdev, A.V. Chubukov and A. Sokol, Phys. Rev. B, {\bf 51}, 14874 (1995). 

\bibitem{ioffe}  L.B. Ioffe and A.J. Millis, Phys. Rev. B, {\bf 51}, 16151 (1995).

\bibitem{qrsg} J. Ye, S. Sachdev, and N. Read, Phys. Rev. Lett. {\bf 70}, 4011
(1993); N. Read, S. Sachdev and J. Ye,
Phys. Rev. B, {\bf 52}, 384 (1995).

\bibitem{oppermann} S. Sachdev, N. Read and R. Oppermann, Phys. Rev. B, {\bf 52}, 10286 (1995).

\bibitem{anirvan} A.M. Sengupta and A. Georges, Phys. Rev. B, {\bf 52}, 10295 (1995).

\bibitem{statphys} S. Sachdev in the {\em Proceedings of the 19th IUPAP
International Conference on Statistical Physics, Xiamen, China, July 31 - August 4 1995\/},
World Scientific, Singapore (1996); Report No.cond-mat/9508080.

\bibitem{oconnor} D. O'Connor and C.R. Stephens, Nucl. Phys. {\bf B360}, 297 (1991);
Proc. Roy. Soc. Lond. A {\bf 444}, 287 (1994); Int. J. Mod. Phys. A {\bf 9}, 2805 (1994);
F. Freire, D. O'Connor, and C.R. Stephens, J. Stat. Phys. {\bf 74}, 219 (1994) and Report No.
cond-mat/9503110.

\bibitem{bgz} E. Brezin, J.C. Le Guillou and J. Zinn-Justin in
{\em Phase Transitions and Critical Phenomena \/}, vol. 6, C. Domb and M.S. Green eds.,
Academic Press, London (1976).

\bibitem{zj} {\em Quantum Field Theory and Critical Phenomena\/} by J. Zinn-Justin,
Oxford University Press, Oxford (1993).

\bibitem{halphoh} P.C. Hohenberg and B.I. Halperin, Rev. Mod. Phys.
{\bf 49}, 435 (1977).

\bibitem{ssising} S. Sachdev, Nucl. Phys. B {\bf 464}, 576 (1996); K. Damle and S.
Sachdev, Phys. Rev. Lett. {\bf 76}, 4412 (1996).

\bibitem{luscher} M. Luscher, Phys. Lett. {\bf 118B}, 391 (1982).

\bibitem{bz} E. Brezin and J. Zinn-Justin, Nucl. Phys. {\bf B257}, 867 (1985).

\bibitem{rudnick} J. Rudnick, H. Guo and D. Jasnow, J. Stat. Phys. {\bf 41}, 353 (1985).

\bibitem{ramond} {\em Field Theory\/} by P. Ramond, Benjamin Cummings, Reading MA (1981).

\bibitem{ma} {\em Modern Theory of Critical Phenomena \/} by S.-K. Ma, Benjamins Cummings,
Reading MA (1976).

\bibitem{nr} D.R. Nelson and J. Rudnick, Phys. Rev. Lett. {\bf 35}, 178 (1975);
A.D. Bruce and D.J. Wallace, J. Phys. A, {\bf 9}, 1117 (1976).

\bibitem{belitz} T. Vojta, D. Belitz, R. Narayan and T.R. Kirkpatrick, cond-mat/9510146;
T.R. Kirkpatrick and D. Belitz, cond-mat/9601008.

\bibitem{bray} D. O'Connor, C.R. Stephens, and A.J. Bray, cond-mat/9601146.


\end{references}
\end{document}